\documentclass[a4paper,fleqn,usenatbib]{mnras}

\usepackage[T1]{fontenc}
\usepackage{ae,aecompl}

\usepackage{graphicx}	% Including figure files
\usepackage{amsmath}	% Advanced maths commands
\usepackage{amssymb}	% Extra maths symbols

\title[The overmassive black hole in NGC 1277]{The overmassive black hole in NGC 1277: new constraints from molecular gas kinematics}

\author[J. Scharw\"achter et al.]{J. Scharw\"achter$^{1}$\thanks{E-mail: julia.scharwaechter@obspm.fr}, 
F. Combes$^{1}$,
P. Salom\'e$^{1}$,
M. Sun$^{2}$
and
M. Krips$^{3}$
\\
$^{1}$LERMA, Observatoire de Paris, PSL, CNRS, Sorbonne Universit\'es, UPMC, F-75014, Paris, France\\
$^{2}$Physics Department, University of Alabama in Huntsville, Huntsville, AL 35899, USA\\
$^{3}$Institut de Radioastronomie Millim\'etrique (IRAM), 300 Rue de la Piscine, Domaine Universitaire, F-38406 Saint Martin d'H\`eres, France
}

\pubyear{0000}

\begin{document}
\label{firstpage}
\pagerange{\pageref{firstpage}--\pageref{lastpage}}
\maketitle

% Abstract of the paper
\begin{abstract}
We report the detection of CO(1-0) emission from NGC~1277, a lenticular galaxy in the Perseus Cluster.
NGC~1277 has previously been proposed to host
an overmassive black hole (BH) compared to the galaxy bulge luminosity (mass), based on stellar-kinematic measurements. 
The CO(1-0) emission, observed
with the IRAM Plateau de Bure Interferometer (PdBI) using both, a
more compact ($2.9$-arcsec resolution) and a more extended ($1$-arcsec resolution) 
configuration, is likely to originate from the
dust lane encompassing the galaxy
nucleus at a distance of 0.9~arcsec ($\sim 320$~pc). The double-horned CO(1-0) profile found at 
$2.9$-arcsec resolution traces $1.5\times 10^8\ M_\odot$ of molecular 
gas, likely orbiting in the dust lane at $\sim 550\ \mathrm{km\ s^{-1}}$, which suggests a total
enclosed mass of $\sim 2\times 10^{10}\ M_\odot$. At 1-arcsec resolution,
the CO(1-0) emission appears
spatially resolved along the dust lane in east-west direction, though at a low signal-to-noise ratio.
In agreement with the previous stellar-kinematic measurements,
the CO(1-0) kinematics is found to be consistent with an $\sim 1.7\times 10^{10}\ M_\odot$ BH for a stellar mass-to-light ratio of $M/L_V=6.3$, while a less massive BH 
of $\sim 5\times 10^{9}\ M_\odot$ is possible when assuming a larger $M/L_V=10$.
While the molecular gas reservoir may be associated with a low level of star formation activity,
the extended 2.6-mm continuum emission is likely to originate from a weak AGN, possibly characterized by an inverted
radio-to-millimetre spectral energy distribution. Literature radio and X-ray data indicate that the BH 
in NGC~1277 is also overmassive 
with respect to the Fundamental Plane of BH activity. 
\end{abstract}


%%%%%%%%%%%%%%%%%%%%%%%%%%%%%%%%%%%%%%%%%%%%%%%%%%

%%%%%%%%%%%%%%%%% BODY OF PAPER %%%%%%%%%%%%%%%%%%

\section{Introduction}

The discovery of scaling relations between supermassive black holes (BHs) and their host galaxies has been a major 
observational result in the field of galaxy evolution over the last $\sim$15--20 yr.
Scaling relations have been found between the mass of the supermassive BH and the stellar velocity dispersion of the host galaxy bulge ($M_\mathrm{BH}$-$\sigma$ relation), the bulge luminosity ($M_\mathrm{BH}$-$L_\mathrm{sph}$ relation), and bulge mass ($M_\mathrm{BH}$-$M_\mathrm{sph}$ relation),
\citep[e.g.][]{Magorrian:1998aa, Ferrarese:2000aa, Gebhardt:2000aa, Marconi:2003aa,Haring:2004aa, Gultekin:2009aa,Graham:2011aa,Sani:2011aa,McConnell:2013aa}.
These empirical scaling relations have been interpreted as a sign of co-evolution between supermassive BHs and their host
galaxies, mediated by AGN feedback \citep[e.g.][]{Silk:1998aa,Di-Matteo:2005aa}. However, the implicit build-up of scaling relations in the course of hierarchical merging, as 
proposed by \citet{Peng:2007aa} and \citet{Jahnke:2011aa},
may play a role as well. A more complex picture involving self-regulatory and non-causal processes may be required to
explain the various aspects of the observed BH-galaxy correlations \citep[see][for a recent review]{Kormendy:2013aa}.

Outliers from the BH scaling relations are of particular interest for the purpose of testing and interpreting
these relations.
Larger sample sizes of galaxies with reliable measurements of BH and galaxy parameters 
have recently led to refined analyses, by which it has been possible to reconcile 
some of the previous outliers with the BH relations. 
Barred galaxies have been found to deviate from the $M_\mathrm{BH}$-$\sigma$ relation of non-barred galaxies, explaining some of the
outliers with under-massive BHs \citep[e.g.][]{Graham:2008aa}.
Similarly, based on a classification into `classical' bulges (formed by mergers) and `pseudo'-bulges (dominated by secular evolution and related to bars), 
\citet{Kormendy:2011aa} have identified `pseudo'-bulges as outliers from the relations.
\citet{Graham:2012aa} and \citet{Graham:2013aa} have recently proposed that the location of galaxies in the $M_\mathrm{BH}$-$M_\mathrm{sph}$ plane can
be interpreted as a bent relation, resulting from the superposition of two different relations for galaxies with `S\'ersic' and `core-S\'ersic' spheroids in the low- and high-mass
regime, respectively. `Core-S\'ersic' spheroids, which are typically more massive and thought to be dominated by dry mergers, 
follow a nearly log-linear $M_\mathrm{BH}$-$M_\mathrm{sph}$ relation, while
 `S\'ersic' spheroids, which are thought to be dominated by rapid BH growth in gas-rich processes, follow a steeper, almost quadratic relation 
 at BH masses below $\sim (2-10)\times 10^8\ M_\odot$.
Some low-redshift AGN with apparently under-massive BHs \citep[e.g.][]{Mathur:2012aa, Busch:2014aa} could be representatives of this near-quadratic relation
\citep{Graham:2015aa}.

At the high-mass end, the BH scaling relations have recently been challenged by a number of galaxies hosting
ultramassive BHs of the order of $10^{10} M_\odot$, the largest BH masses known to date. 
Ultramassive BHs are likely to provide new insights
into galaxy and BH growth, as they must have experienced the strongest growth over time. Ultramassive BHs seem to typically be associated with 
central massive galaxies in clusters. Two such galaxies, NGC~4889 and NGC~3842, have measured BH masses of $2.1^{+1.6}_{-1.6}\times 10^{10}\ M_\odot$ 
and $9.7^{+3.0}_{-2.5}\times 10^9\ M_\odot$, respectively 
\citep{McConnell:2011aa, McConnell:2012aa} and further indirect arguments suggest a similar association \citep{Hlavacek-Larrondo:2012aa}. 
Ultramassive BHs seem to
show a tendency to be overmassive compared to the $M_\mathrm{BH}$-$\sigma$ relation, while they are still largely in agreement
with the $M_\mathrm{BH}$-$M_\mathrm{sph}$ relation \citep[cf.][]{McConnell:2013aa}.
This deviation could indicate that the growth of the central galaxies in clusters has had a significant contribution from dry mergers,
which tend to increase velocity dispersion more than stellar mass \citep{Volonteri:2013aa}. However, the dry-merger scenario is observationally controversial
 \citep{Savorgnan:2015aa}.

An extreme case for an ultramassive BH is NGC~1277, which is not a Brightest Cluster Galaxy, but 
a $1.2\times 10^{11}\ M_\odot$ compact S0 galaxy in the core of the Perseus Cluster, about 3.8~arcmin away from the
Brightest Cluster Galaxy NGC~1275 \citep{van-den-Bosch:2012aa}. 
A number of recent studies have analysed the BH mass in NGC~1277 based on stellar kinematics.
The first measurement was published by \citet{van-den-Bosch:2012aa}, who found a BH mass of $(1.7 \pm 0.3) \times10^{10}\ M_\odot$, using Schwarzschild models
for the long-slit stellar kinematics and input
photometry from multi-Gaussian expansion of a {\it Hubble Space Telescope (HST)} image. 
The measurement by \citet{Yildirim:2015aa}
has resulted in a marginally smaller best-fitting BH mass of 
$1.3 \times10^{10}\ M_\odot$, but, otherwise, in a good general agreement with the \citet{van-den-Bosch:2012aa} results. Most recently,
\citet{Walsh:2015aa} have presented a revised BH mass of
$(4.9\pm1.6) \times10^{9}\ M_\odot$, based on stellar-kinematic data at higher spatial resolution than used in the previous studies.
\citet{van-den-Bosch:2012aa} report an extreme BH-to-stellar mass ratio, with a BH mass amounting to 59 per cent of the
bulge mass or 14 per cent of the total stellar mass of NGC~1277.
Even with the latest BH mass measurement of $(4.9\pm1.6) \times10^{9}\ M_\odot$, 
the BH in 
NGC~1277 is overmassive compared to the $K$-band
$M_\mathrm{BH}$-$L_\mathrm{sph}$ relation \citep{Walsh:2015aa}.

NGC~1277 could have lost a fraction of its stellar mass 
by tidal stripping in the cluster environment, which could explain why the
BH is overmassive.
However, \citet{van-den-Bosch:2012aa} find no strong indications that 
would suggest that NGC~1277 has been affected by tidal stripping.
Optical spectra suggest that the stellar population of NGC~1277 is uniformly old ($\sim 10$~Gyr), 
and the galaxy shows no obvious signs of interactions \citep{van-den-Bosch:2012aa, Trujillo:2014aa, Martin-Navarro:2015aa}.
NGC~1277 has been discussed as a local example of a `relic' galaxy, which has not undergone any significant transformations in its recent
evolution and shows a stellar mass density profile similar to the ones found for massive compact high-redshift galaxies 
\citep{Trujillo:2014aa, Ferre-Mateu:2015aa}.

The overmassive BH in NGC~1277 has possibly been in place since this last violent star formation episode, because
significant BH accretion without coeval star formation is unlikely \citep{van-den-Bosch:2012aa}. 
By analysing an X-ray {\it Chandra} spectrum of NGC~1277, \citet{Fabian:2013aa} find evidence of a
power-law from an unresolved central point source with a 0.5--7~keV luminosity of $1.3\times 10^{40}\ \mathrm{erg\ s^{-1}}$ (corrected for absorption) and
extended thermal gas. 
NGC~1277 has also been reported to be associated with an unresolved radio continuum detection with fluxes of 2.85 and 1.6~mJy
at 1.4 and 5~GHz, respectively \citep{Sijbring:1993}.
\citet{Fabian:2013aa} argue that the current growth rate of the BH is likely to be negligible and that
radiatively efficient Bondi accretion on to such a massive BH would suggest a 5--6 orders of magnitude larger luminosity
of the X-ray power-law source than observed. The small current growth rate is surprising because the 
X-ray properties of NGC~1277 are characteristic of minicoronae 
and the corona lies within the Bondi radius \citep{Sun:2007aa, Fabian:2013aa}.

In view of the extreme properties of NGC~1277, 
\citet{Fabian:2013aa} have proposed an evolutionary scenario based on the
assumption that ultramassive BHs acquired the bulk of their mass at high redshift, where the most massive haloes could grow BHs rapidly. 
This rapid BH growth would most likely have involved extreme AGN feedback, shutting down 
new star formation. By $z= 3$, ultramassive BHs are then likely to be hosted
in compact red bulges with a passive stellar population. \citet{Fabian:2013aa} suggest that the further evolution of these 
host bulges depends on whether 
they are able to accrete external gas and stars. 
Galaxies like NGC~1277, moving at high velocity in a cluster core, may have remained virtually unchanged since $z=3$, 
whilst Brightest Cluster Galaxies, centred in 
the cluster potential well, may have continued to grow significantly.
\citet{Graham:2015aa} point out that the location of NGC~1277 in the
$M_\mathrm{BH}$-$M_\mathrm{sph}$ plane overlaps with an extrapolation of the near-quadratic 
relation for `S\'ersic' spheroids to high masses, where the plane is typically populated by `core-S\'ersic' spheroids.
In this interpretation, NGC~1277 could be a rare example of a galaxy in which growth by gas-rich processes has continued on a near-quadratic relation
instead of being terminated at the typical break mass. This scenario, however, is more difficult to reconcile with the
small present growth rate of the BH in NGC~1277 and the predominantly old stellar population.

The first BH mass measurement for NGC~1277 by \citet{van-den-Bosch:2012aa}, which resulted in
$(1.7 \pm 0.3) \times10^{10}\ M_\odot$, 
has been controversial.
\citet{Emsellem:2013aa} argues that this large value for the BH mass is mostly determined by the fit to the central part of the measured
Gauss-Hermite $h_4$ profile, so that the accuracy of this parameter is critical.
The $N$-body realizations presented by \citet{Emsellem:2013aa} show that 
a smaller BH mass of $5 \times10^9\ M_\odot$ provides a good fit to the stellar kinematics, except for 
a small discrepancy in the Gauss-Hermite $h_4$ parameter.
According to \citet{Emsellem:2013aa}, an even smaller BH mass is possible,
if variations in the mass-to-light ratio or high-velocity stars in the central region are present.
These $N$-body 
realizations are based on a new multi-Gaussian expansion of the {\it HST} photometry
and on the assumption that the potential is defined by a constant mass-to-light ratio of $M/L_V = 10$ without any dark matter halo.
As an alternative scenario, 
\citet{Emsellem:2013aa} also demonstrates that an end-on view on an inner bar together with a BH of 
only $2.5 \times10^9\ M_\odot$
can reproduce the stellar kinematics, except for the central part of the $h_4$ profile.
The smaller BH mass of $5 \times10^9\ M_\odot$ suggested by 
\citet{Emsellem:2013aa} is consistent with the most recent measurement of $(4.9\pm1.6) \times10^{9}\ M_\odot$ reported by
\citet{Walsh:2015aa} based on stellar kinematic data at high spatial resolution.

An independent constraint on the BH mass of NGC~1277 can be obtained by using gas kinematics as
a tracer of the gravitational potential \citep[e.g.][]{Davis:2013aa, Davis:2014aa, Onishi:2015aa}. 
In this paper, we report the first detection of CO(1-0) emission from the centre of NGC~1277, based on observations with
the IRAM PdBI. 
We interpret the CO(1-0) and 2.6-mm continuum data with the help of an optical image,
obtained from archival {\it HST} data,
and discuss implications for the BH mass of NGC~1277.
The observations and data reduction of the IRAM and ancillary data are described in Section~\ref{sec:obs}. The results obtained for the 
CO(1-0) and 2.6-mm continuum emission are presented in Section~\ref{sec:results}.
In Section~\ref{sec:discussion}, we interpret the origin of the molecular gas and continuum detections, derive constraints for the enclosed mass from the 
molecular gas kinematics,
and discuss star formation in the molecular gas as well as BH accretion scenarios. 
We summarize the main results in Section~\ref{sec:summary} and conclude with notes on the
role of NGC~1277 in BH and galaxy growth scenarios.
Throughout the paper, we will assume a scale of 353~pc~arcsec$^{-1}$ and a luminosity distance modulus
of 34.39 (i.e. a luminosity distance of 75.5~Mpc), consistent with the values used by \citet{van-den-Bosch:2012aa}.

\section{Observations and data reduction}\label{sec:obs}

\subsection{IRAM PdBI data}

The millimetre data for NGC~1277 were obtained using the 
IRAM PdBI in both, a more extended and a more compact configuration, providing spatial resolutions of 
$\sim 1$~arcsec and $\sim 2.9$~arcsec, respectively. A list of all observations, indicating the number of antennas, the 
minimum and maximum baselines, and the calibrators, is provided in Table~\ref{tab:obs}.
In order to observe the redshifted CO(1-0) line in NGC~1277, the dual-polarization receiver in the 3-mm band was tuned to 113.356~GHz, according
to the optical systemic velocity of NGC~1277 of $5066\ \mathrm{km\ s^{-1}}$ \citep{Falco:1999aa}. 
We verified the systemic velocity based on stellar absorption lines 
(\ion{Ca}{i}~$\lambda 2.26\ \mu$m, \ion{Mg}{i}~$\lambda 2.28\ \mu$m, and CO bandheads at about 2.29, 2.32, and 2.35~$\mu$m)
in a single nuclear-aperture $K$-band spectrum of NGC~1277,
extracted from a subset of archival NIFS (Gemini North) data from programme GN-2011B-Q-27 (PI: D. Richstone), \citep[see ][]{Walsh:2015aa}.
The systemic velocity derived from fitting this spectrum with template stars from the NIFS spectral template library v2.0 \citep{Winge:2009aa}
using {\sc pPXF} \citep{Cappellari:2004aa} agrees within $\sim 20\ \mathrm{km\ s^{-1}}$ with the value of $5066\ \mathrm{km\ s^{-1}}$. 
Since an offset of this order of magnitude is negligible compared to the accuracy at which the CO(1-0) data are analysed,
we will present the CO(1-0) data using the reference velocity of $5066\ \mathrm{km\ s^{-1}}$.

The PdBI observations of NGC~1277 were performed using the Wide-Band correlator WideX, which provides a spectral resolution of 1.95~MHz and a bandwidth of 3.6~GHz covering the
CO(1-0) line as well as the adjacent 2.6-mm continuum. The J2000 coordinates
$\mathrm{R.A.}=03$h~19m~51.5s and $\mathrm{Dec}=+41^\circ$~34\arcmin~24\farcs7 were used as pointing reference. 
The data for both configurations were reduced using the IRAM {\sc GILDAS} software {\sc CLIC} and {\sc MAPPING}.
\begin{table*}
 \centering
 \begin{minipage}[c]{165mm}
  \caption{Journal of observations}
  \label{tab:obs}
  \begin{tabular}{lccccccc}
  \hline
   Configuration & Observation     &   Number of & Minimum & Maximum &                          RF & Phase/amplitude & Flux    \\
              &                 date        &                     antennas                & baseline (m) & baseline (m)                  &  calibrator & calibrator & calibrator \\
  \hline
 Compact & 2014-Dec-16 &  6 & 24 & 176 & 3C~454.3 & 3C~84, 0300+470 & MWC~349 \\
                & 2015-Feb-19 & 7 & 24 & 144 & 3C~84       & 3C~84, 0300+470 & MWC~349 \\ 
                & 2015-Apr-11 & 7 & 24 & 176 & 0059+581  & 3C~84, 0300+470 & LKHA~101 \\
                & 2015-Jul-03  & 6 & 24 & 97 & 3C~454.3  & 3C~84, 0300+470 & MWC~349 \\
                & 2015-Jul-06  & 6 & 24 & 97 & 3C~454.3  & 3C~84, 0300+470 & MWC~349 \\
  \hline
 Extended & 2014-Mar-12 & 6 & 88 & 452 & 3C~273 & 3C~84, 0307+380 & 0307+380 \\
                 & 2014-Mar-14 & 6 & 88 & 452 & 2013+370 & 3C~84, 0307+380 & MWC~349 \\
\hline
\end{tabular}
\end{minipage}
\end{table*}

\subsubsection{Observations at low spatial resolution}\label{sec:reduceCD}
The observations in the more compact configuration (special CD configuration, including the new seventh antenna) -- obtained 
between 2014 December 16 and 2015 July 6 -- cover
baselines between 24 and 176~m and a total on-source time of 9.38~h for seven antennas. The details
of these observations are listed in Table~\ref{tab:obs}.
The phase and amplitude calibration was performed based on 3C~84 and quasar 0300+470. 
Either MWC~349 or LKHA~101 were used as flux calibrators, while 
0059+581, 3C~454.3, or 3C~84 were used as RF calibrators. The 
RF calibration based on 3C~84 could be affected by CO(1-0) emission from 3C~84. This calibrator was nevertheless chosen 
for the observations on 2015 February 19, because no obvious contamination was found and because the calibration was improved by
using this strong calibrator instead of the weaker alternatives. 
The final UV table was created by merging the data from all observing dates using a spectral resolution of 40~MHz
($105.8\ \mathrm{km\ s^{-1}}$).

The UV table was mapped using $256\times 256$
spatial pixels, 
natural weighting, no tapering, and a pixel size of $0.61\times 0.61$~arcsec$^{2}$. 
The data cube was cleaned by applying H\"ogbom deconvolution
\citep{Hogbom:1974aa} to each spatial plane. The deconvolution was performed with a predefined support covering the emission that is evident in each
channel close to the
reference position of NGC~1277.
The resulting beam size is $2.96 \times 2.78$~arcsec$^2$.
The 1$\sigma$ noise level per 40-MHz channel in the cleaned cube is found to be 0.36~mJy~beam$^{-1}$ at $6000\ \mathrm{km\ s^{-1}}$ and increases to 
0.62~mJy~beam$^{-1}$ at $-3400\ \mathrm{km\ s^{-1}}$ with a value of about
0.47~mJy~beam$^{-1}$ around the NGC 1277 reference velocity. 
The continuum was subtracted from the cleaned data cube by subtracting the average of all channels close to the line emission but 
excluding the velocity range of
the line itself.
A number of the highest and lowest velocity channels were not considered when computing the continuum, since they show indications
of unidentified features or noise peaks.

\subsubsection{Observations at high spatial resolution}\label{sec:reduceB}

In the more extended configuration (B configuration), NGC 1277 was observed 
on 2014 March 12 and 14, respectively (see Table~\ref{tab:obs}). These observations cover baselines
between 88 and
452~m and a total on-source time of 8.29~h with six antennas.
The phase and amplitude calibration 
for both observing dates was derived from observations of 3C~84 and quasar 0307+380. The flux is based on
0307+380 or MWC~349, while 3C~273 or 2013+370 were used as RF calibrators.
The final UV table was created by merging the data from both observing dates using 
a spectral resolution of 40~MHz or 105.8~km~s$^{-1}$.

The UV table was converted into a data cube by mapping the data spatially on to $512\times 512$
pixels using 
natural weighting, no tapering, and a pixel size of $0.24\times 0.24$~arcsec$^{2}$. 
The data cube was cleaned using the same method as applied to the low-resolution data.
 The resulting beam size in this more extended configuration is $1.15 \times 0.86$~arcsec$^2$.
The 1$\sigma$ noise level per 40-MHz channel in the cleaned cube increases with decreasing
velocity from about 0.3~mJy~beam$^{-1}$ at $6000\ \mathrm{km\ s^{-1}}$ to 0.6~mJy~beam$^{-1}$ at $-3400\ \mathrm{km\ s^{-1}}$ with a value of
0.42~mJy~beam$^{-1}$ around the NGC 1277 reference velocity.

Since the CO(1-0) emission in this data set is weak, 
a blind detection assessment was performed on the cleaned data cube, both before and after continuum subtraction, using a wrapper
around {\sc SExtractor} \citep{Bertin:1996aa}. For each channel in the cube, a narrow-band image was created by 
averaging over a predefined number of neighbouring channels. In order to account for different line widths, 
the channel average was varied between a single channel and the 
maximum range of channels, using only odd numbers for a symmetric average around the central channel.
Each narrow-band image was sent to {\sc SExtractor} in order to identify all groups of pixels with values above a certain detection threshold. 
Both, the detection threshold as well as the minimum area covered by pixels above this threshold, are user-defined input parameters of {\sc SExtractor}.
In order to improve the detection process, we applied the {\sc SExtractor} image filtering before detection, using a 
Gaussian convolution kernel with an FWHM of 4~pixels (0.96~arcsec), which roughly matches the beam size.

The detection assessment of the full data cube using single-channel narrow-band images, a {\sc SExtractor} detection threshold of 3$\sigma$, and a 
minimum detection area of 3 pixels, results in the detection of a continuum emission peak 
at a mean offset of $\Delta\mathrm{R.A.}=-0.15$~arcsec and $\Delta\mathrm{Dec}=0.18$~arcsec for most channels across the full frequency range.
In order to isolate weak line emission, a continuum subtraction was performed using the same method
as applied to the low-resolution data. The resulting continuum image has a 1$\sigma$ background noise level of 0.089 mJy~beam$^{-1}$.
A new detection assessment was performed on the continuum-subtracted data cube, using a detection threshold of 2$\sigma$ and a minimum detection area of 1 pixel
to detect faint features. Within a distance of 13~arcsec around the reference position, only 
one detection is identified at a $>5\sigma$ (peak flux over background rms) level.
This detection at $\sim 5.1\sigma$ is found at  
$\Delta\mathrm{R.A.}=0.48$~arcsec and $\Delta\mathrm{Dec}=0.17$~arcsec from the reference position when averaging
3 channels ($320\ \mathrm{km\ s^{-1}}$) around a central velocity of 
 $+530\ \mathrm{km\ s^{-1}}$. While this detection is rather marginal based on the blind assessment, we will show in 
 Sect.~\ref{sec:CO} that this feature is clearly confirmed as
 CO(1-0) emission from NGC~1277 by comparison with the data obtained at lower spatial resolution.

\subsection{Ancillary optical data}\label{sec:reduceHST}

Ancillary archival data from {\it HST} observations in the $F550M$ filter 
are used to compare the millimetre observations with the optical features of NGC~1277.
The optical {\it HST} image has already been presented by other authors to discuss the morphology
of NGC~1277
\citep{van-den-Bosch:2012aa,Emsellem:2013aa,Fabian:2013aa,Trujillo:2014aa}.
We found the World Coordinate System (WCS) of the archival {\it HST} image to be offset,
which we corrected by comparing the positions of objects in a $\sim 45 \times 45$~arcsec$^2$ cut-out around NGC~1277 to 
coordinates from the Guide Star Catalog (GSC) 2.3 \citep{Lasker:2008aa}. Four of these objects, extracted using the 
IMSTAR routine included in the {\sc WCSTOOLS} package \citep[e.g.][]{Mink:1997aa}, were found to match catalogued sources.
Based on these, a new plate solution was derived using the {\sc IRAF}\footnote{{\sc IRAF} is distributed by the National Optical Astronomy Observatories, which are operated by the Association of Universities for Research in Astronomy, Inc., under cooperative agreement with the National Science Foundation.} task {\sc CCMAP} with a second-order polynomial fit. The fit shows an rms of about 0.2~arcsec, which is a similar order of magnitude as the average absolute
astrometry error of 0.2--0.3~arcsec of the GSC 2.3 \citep{Lasker:2008aa}.
The {\it HST} image was resampled according to the new plate solution on to $0.05 \times 0.05$~arcsec$^2$ pixels using the 
software {\sc SWarp} \citep{Bertin:2002aa} with a
LANCZOS3 interpolation and an oversampling factor of 3. The final image quality, based on stars in the image field-of-view, is 
$\sim 0.14$--$0.15$~arcsec (Gaussian FWHM).

\section{Results}\label{sec:results}

\subsection{Optical morphology}\label{sec:sysvel}
\begin{figure}
\centering
\includegraphics[width=\linewidth]{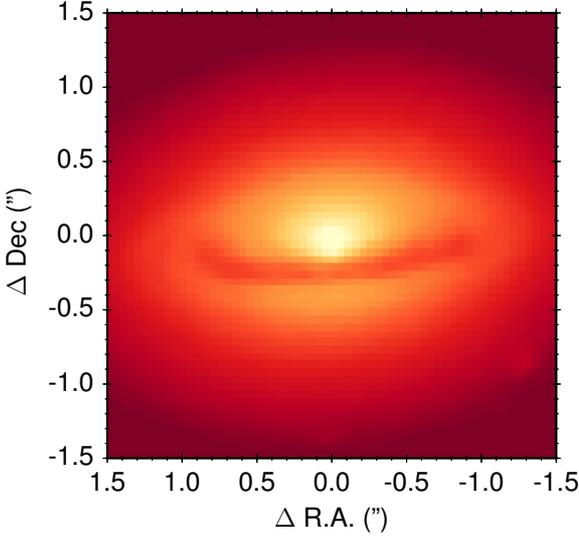}
\caption{Optical $F550M$ {\it HST} image of the central $3 \times 3$~arcsec$^2$ region of NGC~1277. The  
image is shown in arbitrary units in a logarithmic colour scale chosen to enhance the contrast on the dust lane. 
The centre of the relative coordinate system is determined from  
a Gaussian fit to the nuclear emission peak. 
\label{fig:hst}}
\end{figure}

As already discussed by other authors \citep{van-den-Bosch:2012aa,Emsellem:2013aa,Fabian:2013aa}, the {\it HST} optical image shows a prominent dust lane
around the centre of NGC~1277 (Fig.~\ref{fig:hst}).
The dust lane has a radius of $\sim 0.9$~arcsec ($\sim 320$~pc), \citep[see also, e.g.,][]{Fabian:2013aa}, and is seen nearly edge-on.
\citet{van-den-Bosch:2012aa} deduce an inclination of $\sim 75^\circ$, suggesting
a similar inclination for the galaxy, if the dust lane is in the plane of the galaxy disc. 
It has been noted by 
\citet{Fabian:2013aa} that the large central mass concentration corresponding to an $\sim 10^{10}\ M_\odot$ BH in NGC~1277 would imply
high circular velocities of $480\ \mathrm{km\ s^{-1}}$ for gas orbiting in the dust lane.

\subsection{2.6-mm continuum emission}\label{sec:cont}
\begin{figure}
\centering
\includegraphics[width=\linewidth]{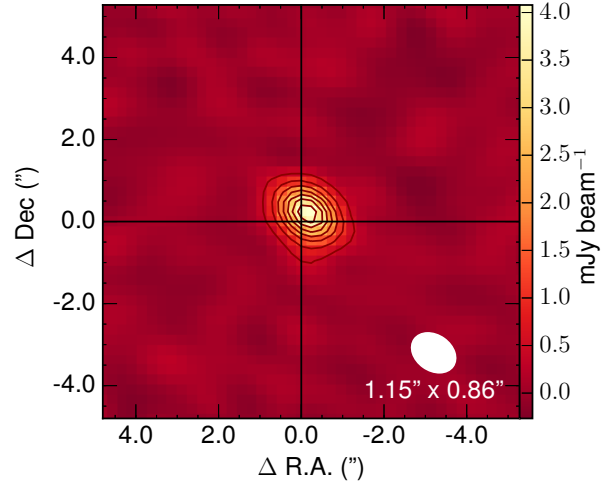}
\caption{Map of the 2.6-mm continuum of NGC~1277 based on the data obtained in the more extended configuration. 
Contour levels start at 5$\sigma$ and are shown in steps of 6$\sigma$
(at 0.44, 0.98, 1.51, 2.04, 2.58, 3.11, 3.64~mJy~beam$^{-1}$).
The colour scale extends from $-3\sigma$ to the continuum peak value. 
The size and orientation of the beam are indicated in the lower right corner. \label{f:cont}}
\end{figure}

\begin{figure}
\centering
\includegraphics[width=\linewidth]{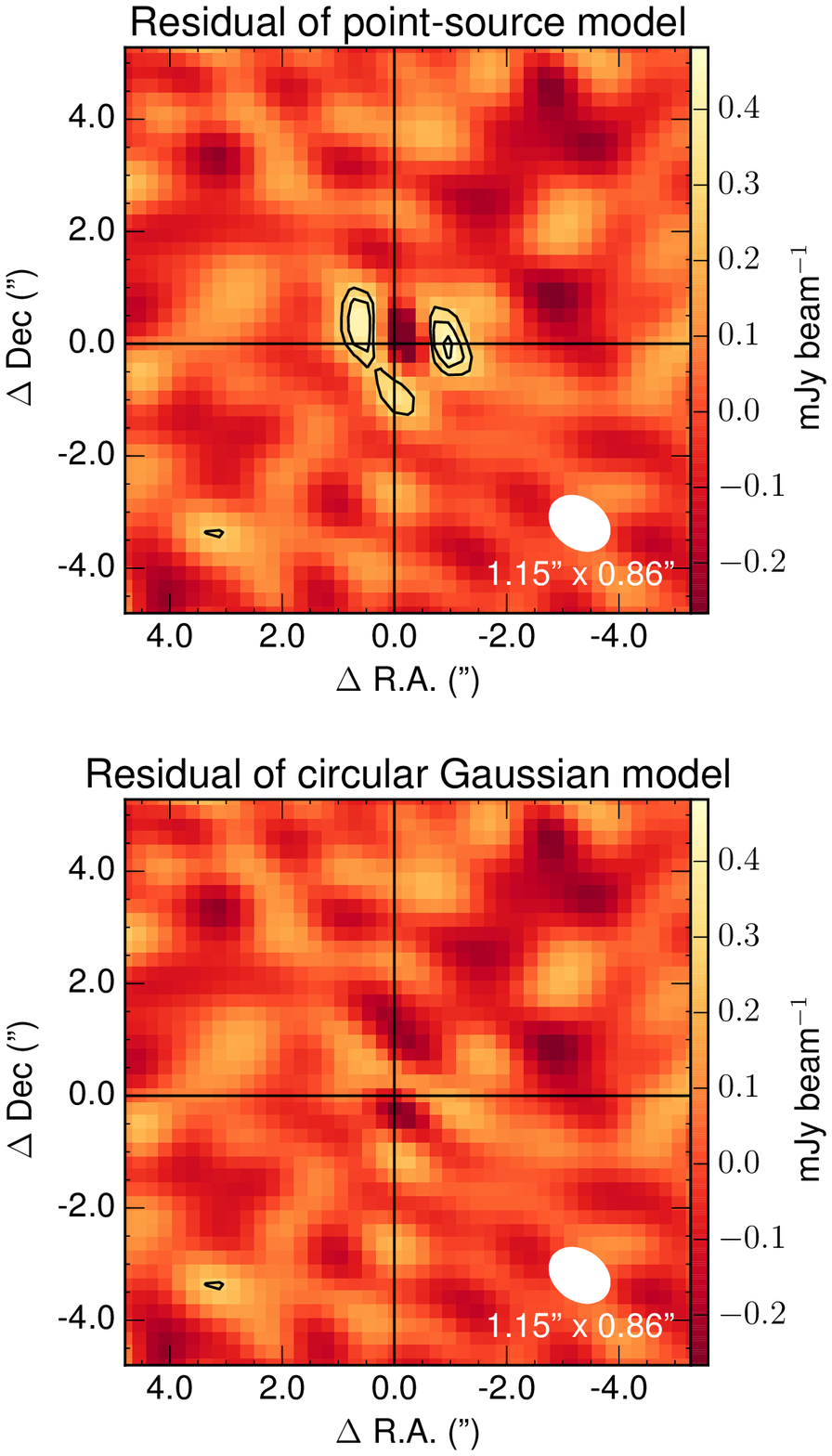}
\caption{Residual maps after subtracting a point source model (top) and a circular
Gaussian model (bottom) from the 2.6-mm continuum of NGC~1277, based on the high-resolution data. 
The 1$\sigma$ noise level of the maps is 0.089~mJy~beam$^{-1}$.
Contour levels start at 3$\sigma$ and are shown in steps of 1$\sigma$
(at 0.27, 0.35, and 0.44~mJy~beam$^{-1}$).
The colour scale in both plots is identical and ranges between $-3\sigma$ and the peak value of the
map in the top panel. 
The size and orientation of the beam are indicated in the lower right corner. \label{f:cont-resid}}
\end{figure}

The 2.6-mm continuum of NGC~1277 (Fig.~\ref{f:cont}), 
obtained from the high-spatial-resolution data,
shows an integrated flux of 
$(5.6 \pm 0.2)$~mJy in a pseudo-circular aperture with a diameter of 3.36~arcsec.
The continuum is found to be extended at the spatial resolution of $1$~arcsec and shows a possible weak
feature to the south.
In order to probe the continuum extension, both, a point source and a circular Gaussian model, were fitted to the continuum map
after convolution of the models with the beam. 
Both models result in
$\Delta\mathrm{R.A.}=-0.15$~arcsec and $\Delta\mathrm{Dec}=0.18$--0.19~arcsec for the continuum position 
with respect to the pointing reference. This is in agreement
with the position derived from the basic blind detection assessment in Sect.~\ref{sec:reduceB}.
Fitting a point source model, yields a best-fitting flux of $4.7$~mJy. It is obvious that this fit only recovers
about 80~per~cent of the total flux of 5.6~mJy, obtained by direct aperture integration.
The residual map of the continuum after subtraction of the point-source model (Fig.~\ref{f:cont-resid}, top panel) shows that
the continuum has extended excess emission compared to a point source.
Fitting a single circular Gaussian instead of a point source, yields best-fitting parameters of 
$5.5$~mJy for the flux and $0.5$~arcsec for the Gaussian FWHM. The circular Gaussian model 
recovers most of the flux measured via direct aperture integration, and the residuals (Fig.~\ref{f:cont-resid}, bottom panel) 
are negligible.

\subsection{CO(1-0) emission}\label{sec:CO}

\begin{figure}
\centering
\includegraphics[width=\linewidth]{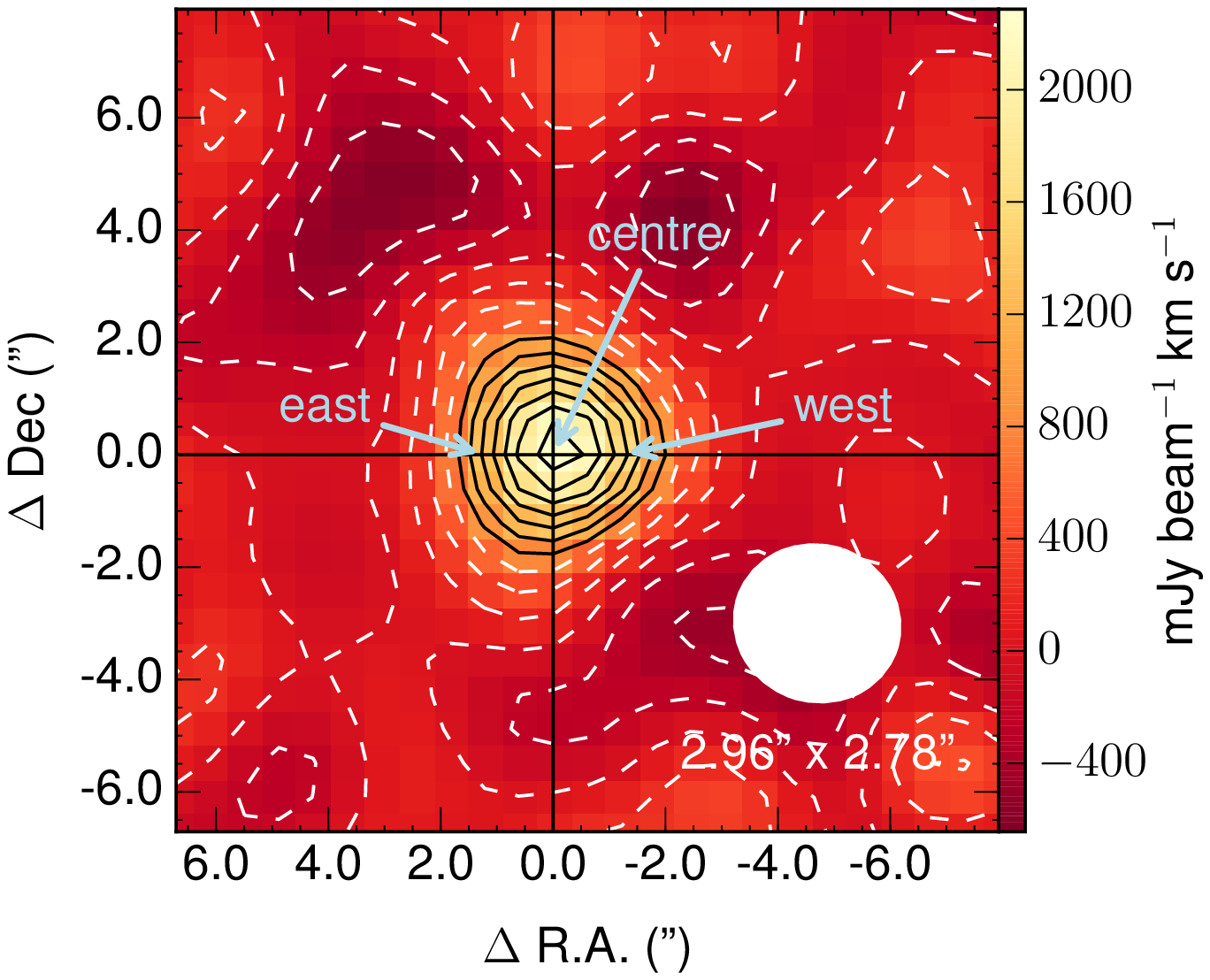}
\caption{Map of the CO(1-0) detection for NGC~1277 obtained in the more compact configuration. The CO(1-0) line has been 
integrated over a symmetric velocity range of 
1380~km~s$^{-1}$ (13 channels of 40~MHz) around the systemic velocity. 
The 4$\sigma$ to 10$\sigma$ contours 
(859, 1074, 1289, 1503, 1718, 1933, and 2148~$\mathrm{mJy\ beam^{-1}\ km\ s^{-1}}$)
are shown as black solid lines. 
The $-3\sigma$ to 3$\sigma$ 
contours 
($-644$, $-430$, $-215$, $0.0$, 215, 430, and 644~$\mathrm{mJy\ beam^{-1}\ km\ s^{-1}}$) are shown as white dashed lines.
The size and orientation of the beam are indicated in the lower right corner. `east', `centre', and `west' mark the
positions of the three pixels from which the marginally resolved spectra shown in Fig.~\ref{f:specCDresolved} were extracted.\label{f:coCD}}
\end{figure}

\begin{figure}
\centering
\includegraphics[width=\linewidth]{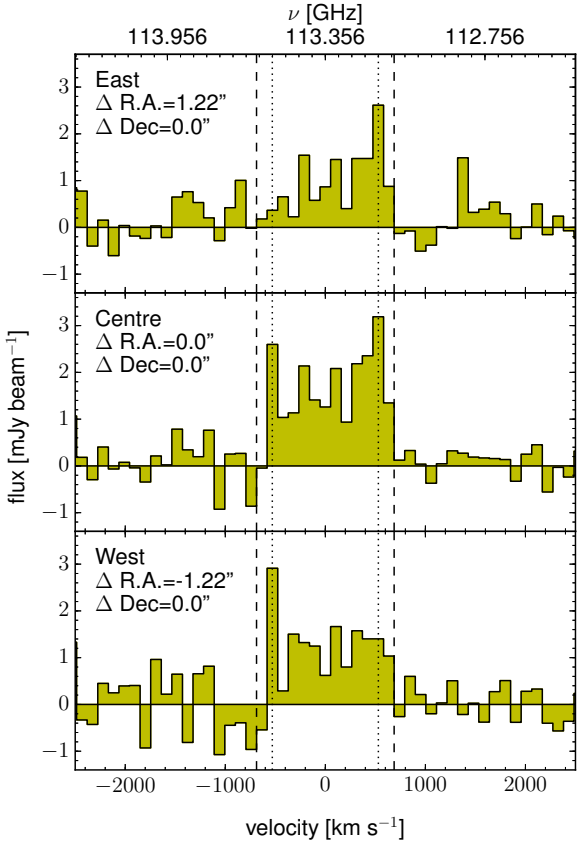}
\caption{Spectra extracted from an eastern, central, and western pixel in the CO(1-0) emission peak,
as indicated in Fig.~\ref{f:coCD}. The spectra are shown at a spectral resolution of 40 MHz (105.8~$\mathrm{km\ s^{-1}}$). 
The vertical dotted lines mark $+530$ and $-530\ \mathrm{km\ s^{-1}}$, 
which roughly coincides with the peaks of the double-horned line profile. The vertical dashed lines indicate the 
symmetric spectral range of 13 channels around the systemic velocity
that was used to compute the integrated CO(1-0) map presented in Fig.~\ref{f:coCD}.\label{f:specCDresolved}}
\end{figure}

\begin{table*}
\centering
 \begin{minipage}[c]{120mm}
  \caption{Summary of the physical properties of the three main CO(1-0) detections, i.e.
  the double-horn profile at 2.9~arcsec spatial resolution as well as the clear and marginal detections of the redshifted and blueshifted horn, respectively, 
  at 1~arcsec resolution. The columns show the integrated flux, the corresponding H$_2$ mass calculated via equations 3 and 4 from \citet{Solomon:2005aa} assuming 
  a Milky Way conversion factor of $\alpha=4.6\ M_\odot\ (\mathrm{K\ km\ s^{-1}\ pc^2})^{-1}$, and the line centre and 
  width, as used for the integrated line maps (see Figs~\ref{f:coCD}--\ref{f:cospec}). }
  \label{tab:co}
  \begin{tabular}{cccccc}
  \hline
  Detection  & Beam size                           & Flux                                       & $M(\mathrm{H_2})$                  & Line centre                    & Line width \\
                                           &  (arcsec$^2)$                  & ($\mathrm{mJy\ km\ s^{-1}}$) & ($10^8\ M_\odot$) & ($\mathrm{km\ s^{-1}}$) & ($\mathrm{km\ s^{-1}}$) \\
\hline
Double-horn profile   &  $2.96 \times 2.78$ & $2400 \pm 400$& $1.5\pm 0.3 $ & 0 & 1380 \\
Redshifted horn        & $1.15 \times  0.86$  & $300 \pm 100$ & $0.19\pm 0.06$ & 530 & 320 \\
Blueshifted horn      & $1.15 \times  0.86$  & $200 \pm 100$ & $0.13\pm 0.06$ & -420 & 320 \\
\hline
\end{tabular}
\end{minipage}
\end{table*}

A very broad CO(1-0) line is clearly detected close to the pointing reference in the 
data obtained at low spatial resolution (Figs~\ref{f:coCD} and \ref{f:specCDresolved}). The line is symmetrically centred around the
systemic velocity of NGC~1277 and marginally resolved in terms of kinematics.
The map of the CO(1-0) emission, shown in Fig~\ref{f:coCD}, has been computed by spectrally 
integrating the data over a symmetric range of 13 channels of 40-MHz width (i.e. $1380\ \mathrm{km\ s^{-1}}$)
around the systemic velocity of NGC~1277. This wide spectral range has been selected to cover the 
full CO(1-0) line profile, as marked by the vertical dashed lines in the middle panel of Fig.~\ref{f:specCDresolved}, where
the spectrum from the brightest pixel of the emission peak is displayed.
The flux of the CO(1-0) emission peak, integrated over a pseudo-circular aperture with a diameter of 6.1 arcsec, 
and the corresponding H$_2$ mass are listed in Table~\ref{tab:co}.

The central CO(1-0) profile (Fig.~\ref{f:specCDresolved}, middle panel)
shows clear indications of a double-horned shape with a stronger velocity component centred at $+530\ \mathrm{km\ s^{-1}}$
and a less pronounced component centred at $-530\ \mathrm{km\ s^{-1}}$. The comparison of the 
three spectra from an 
eastern, central, and western pixel in Fig.~\ref{f:specCDresolved} suggests that the CO(1-0) kinematics is marginally resolved in east-west direction.
The pronounced horn at $+530\ \mathrm{km\ s^{-1}}$ is mostly contributed by the eastern part of the
CO(1-0) emission peak, while the weaker horn at $-530\ \mathrm{km\ s^{-1}}$ seems to be associated with the western part.

\begin{figure}
\centering
\includegraphics[width=\linewidth]{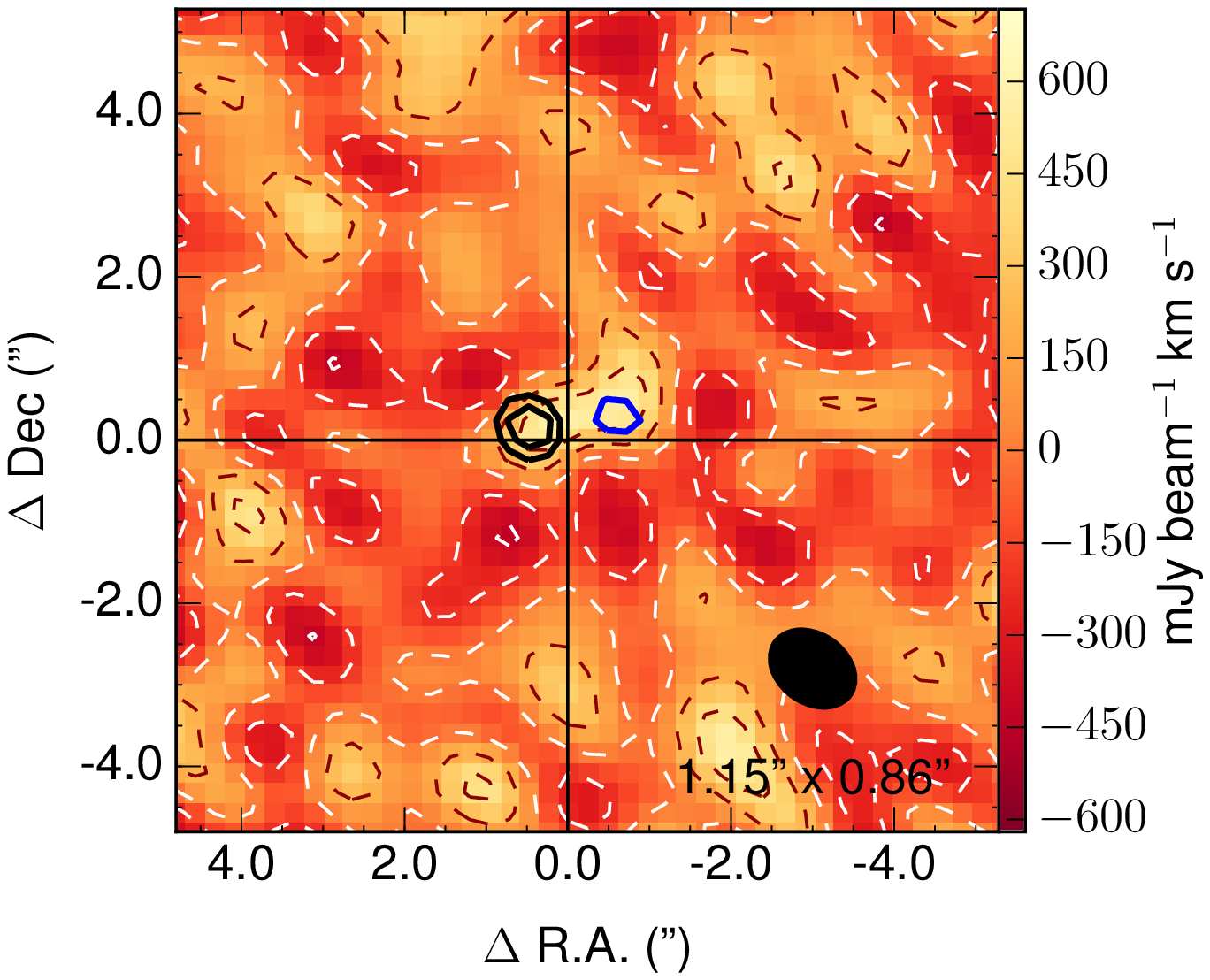}
\caption{Map of CO(1-0) emission features in the more extended configuration. The two (black)
3$\sigma$ and 4$\sigma$ 
contours (232 and 310~$\mathrm{mJy\ beam^{-1}\ km\ s^{-1}}$) east of the reference position
show the only clear CO(1-0) detection,
obtained by spectrally integrating the data cube over three channels of 40~MHz (i.e. $320\ \mathrm{km\ s^{-1}}$) around a central velocity
of $+530\ \mathrm{km\ s^{-1}}$ (see also Sect.~\ref{sec:reduceB}). This detection corresponds to the redshifted horn of the double-horned
profile detected in the more compact configuration (Fig.~\ref{f:specCDresolved}).
The (blue) 3$\sigma$ contour (263~$\mathrm{mJy\ beam^{-1}\ km\ s^{-1}}$) west of the reference position
shows a marginal detection from the blueshifted horn of the profile, obtained by spectrally
integrating
three channels of 40~MHz (i.e. $320\ \mathrm{km\ s^{-1}}$) around a central velocity
of $-420\ \mathrm{km\ s^{-1}}$.
The underlying colour
image shows the map obtained by integrating the cube over the full spectral range of the double-horned profile
(i.e. 13 channels of 40~MHz or $1380\ \mathrm{km\ s^{-1}}$, as indicated by the dashed
vertical lines in Fig.~\ref{f:specCDresolved}). White dashed contours mark the 0 to $-3\sigma$ levels
at 0.0, $-207$, $-413$, and $-620$~$\mathrm{mJy\ beam^{-1}\ km\ s^{-1}}$, respectively. Dark red dashed contours indicate the
1$\sigma$ and 2$\sigma$ levels at 207 and 413~$\mathrm{mJy\ beam^{-1}\ km\ s^{-1}}$, respectively. Note that 3$\sigma$ is not reached. 
The size and orientation of the beam are indicated in the lower right corner.\label{f:co}}
\end{figure}

\begin{figure}
\centering
\includegraphics[width=1.0\linewidth]{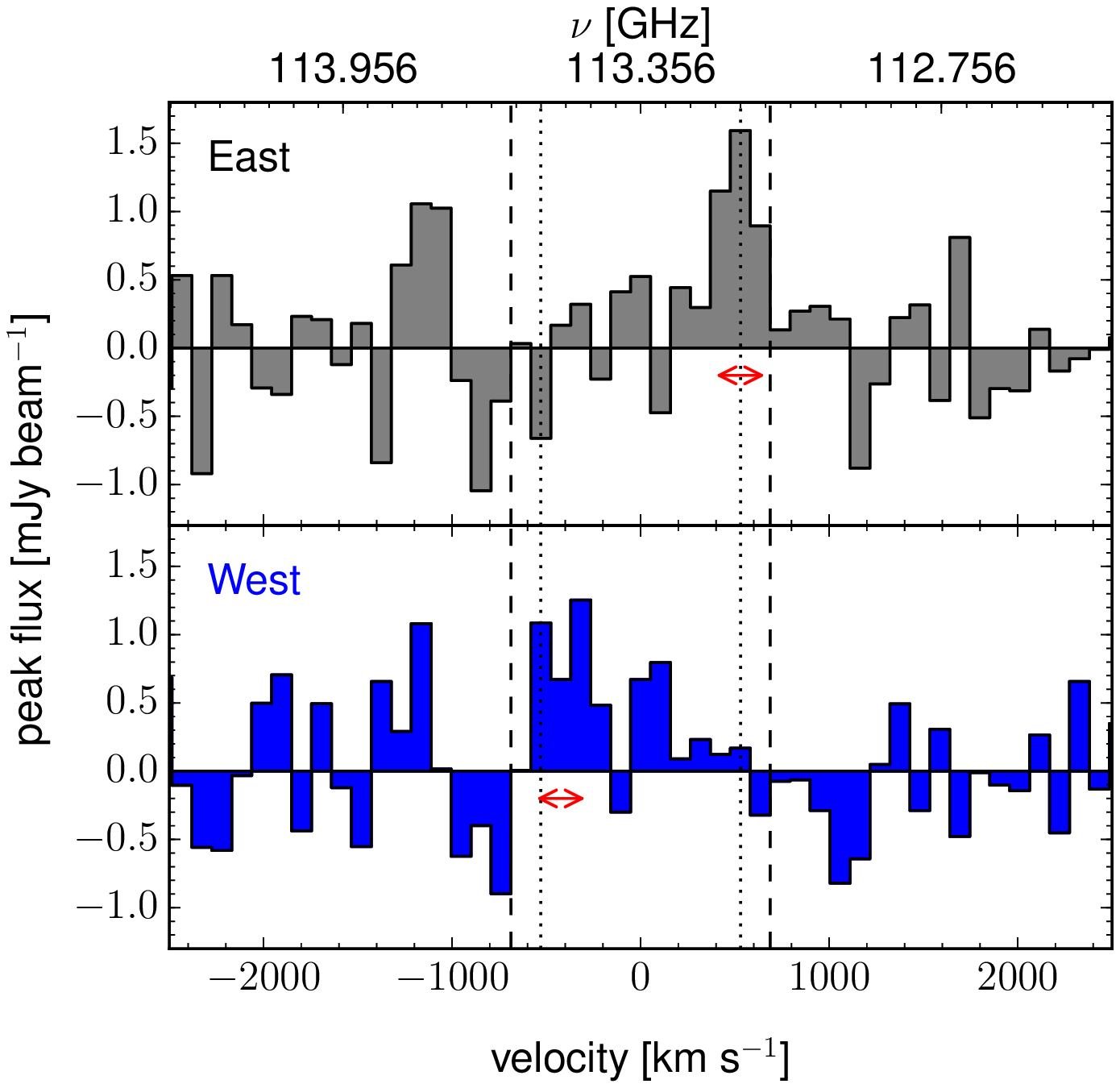}
\caption{Spectra of the CO(1-0) detections in the more extended configuration. 
The upper panel shows the spectrum extracted from the brightest pixel in the eastern emission peak (black contours
in Fig.~\ref{f:co}). The lower panel shows the corresponding spectrum for the western, marginal emission peak (blue contour in Fig.~\ref{f:co}). 
The arrows indicate the spectral integration ranges that were used to compute the maps of the emission peaks
shown as solid (black and blue) contours in Fig.~\ref{f:co}.
For comparison with the double-horn profile seen in the data at low spatial resolution, the vertical dotted and dashed lines from 
Fig.~\ref{f:specCDresolved} are reproduced here.
\label{f:cospec}}
\end{figure}

At 1-arcsec resolution, in the more extended configuration, a clear 
$\sim5\sigma$ detection is only 
found for the pronounced velocity component at $+530\ \mathrm{km\ s^{-1}}$, as deduced from the blind detection assessment in Sect.~\ref{sec:reduceB}.
The narrow-band image of this single CO(1-0) detection is
indicated via the two solid (black) contours in Fig.~\ref{f:co}. This image has been averaged 
over three 40-MHz channels (i.e. $320\ \mathrm{km\ s^{-1}}$) around a central velocity of $+530\ \mathrm{km\ s^{-1}}$
in order to isolate the line emission as suggested by the detection experiment.
The emission peak is found at $\Delta\mathrm{R.A.}=0.48$~arcsec and $\Delta\mathrm{Dec}=0.15$~arcsec, in agreement with the 
position derived from the detection experiment.
The total flux measured for this peak in the above image, using a pseudo-circular aperture with a diameter
of 1.92 arcsec, and the corresponding H$_2$ mass are listed in Table~\ref{tab:co}.
The spectrum of this emission peak, extracted from the brightest pixel, is shown in the top panel of Fig.~\ref{f:cospec}.
The spectrum shows a narrow (roughly three-channel-wide) line centred at $+530\ \mathrm{km\ s^{-1}}$. In comparison to Fig.~\ref{f:specCDresolved},
it is obvious that this single detection corresponds to the pronounced red horn of the 
double-horned CO(1-0) profile seen in the more compact configuration. Fig.~\ref{f:cospec} also 
shows a number of apparent features in the spectral region at $<-1000\ \mathrm{km\ s^{-1}}$, which
we ascribe to the increasing noise towards higher frequencies. 

By visually inspecting the 1-arcsec resolution data in light of
the double-horned profile seen at 2.9-arcsec resolution, we found indications of a marginal CO(1-0) emission peak associated with the blueshifted horn of the profile.
This peak is just above the 3$\sigma$-level when spectrally integrating the cube over $320\ \mathrm{km\ s^{-1}}$ around a central velocity
of $-420\ \mathrm{km\ s^{-1}}$ (single (blue) contour in Fig.~\ref{f:co}). The spectrum of this peak is shown in the lower panel of Fig.~\ref{f:cospec}, and the
corresponding total flux and gas mass, derived by integrating the emission over a pseudo-circular aperture with a diameter
of 1.92 arcsec, are listed in Table~\ref{tab:co}.
The peak is found at $\Delta\mathrm{R.A.}=-0.63$~arcsec and $\Delta\mathrm{Dec}=0.32$~arcsec, i.e. to the west of the detection associated
with the redshifted horn. Although the current signal-to-noise ratio is low, 
the 1-arcsec resolution data seem to spatially resolve the east-west alignment of the redshifted and blueshifted CO(1-0) emission that 
has already been indicated by the data at 2.9-arcsec resolution (Fig.~\ref{f:specCDresolved}).

Possible evidence for CO(1-0) emission from the region connecting the two emission peaks is found when integrating the 1-arcsec resolution data
over the full spectral range covered by the double-horned profile seen at 2.9-arcsec resolution (i.e. $1380\ \mathrm{km\ s^{-1}}$). 
None of the features in the resulting map, shown as the underlying
colour image in Fig.~\ref{f:co}, reaches the 3$\sigma$ limit. However, the map suggests that the two CO(1-0) emission peaks are located at
the opposite ends of an underlying ridge of positive emission. In view of the double-horned profile found at lower spatial resolution, it can be speculated that
this emission ridge originates from components of the
double-horned profile that are just below the current detection limit.

\begin{figure}
\centering
\includegraphics[width=\linewidth]{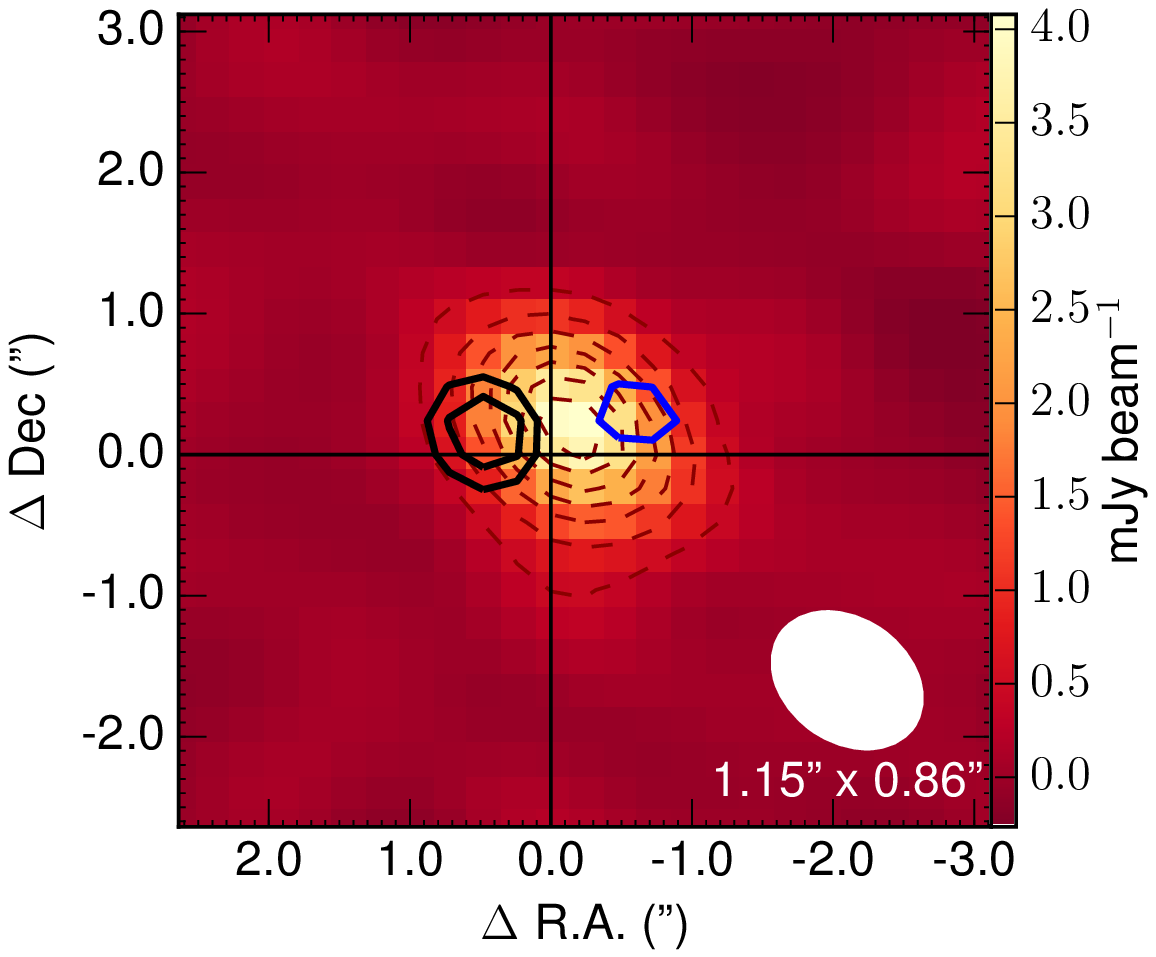}
\caption{Position of the CO(1-0) emission peaks relative to the 2.6-mm continuum emission in the data at 1-arcsec resolution. 
The continuum colour scale and contours (dark red dashed lines) are the same as in Fig.~\ref{f:cont}. 
The contours for the two CO(1-0) peaks (black and blue)
are shown as in Fig.~\ref{f:co}.
The size and orientation of the beam are indicated in the lower right corner.\label{f:cont-co}}
\end{figure}

The clearly detected CO(1-0) emission peak associated with the velocity of $+530\ \mathrm{km\ s^{-1}}$ 
shows a 
spatial offset of 0.6~arcsec to the east compared to the 
2.6-mm continuum emission (Fig.~\ref{f:cont-co}).
Assuming that the $1\sigma$ positional uncertainty $\Delta\Theta$
for a signal-to-noise ratio SNR and a beam size $\Theta_B$ is given by
$\Delta\Theta \approx \Theta_B/(2\times \mathrm{SNR})$, the CO(1-0) line emission involves a 
positional uncertainty of $\Delta \Theta \sim 0.1$~arcsec (using the major axis of the beam size (1.15~arcsec)
and the SNR of the CO(1-0) emission peak of $\sim 5.1$). Compared to this estimate, the 
measured offset between the CO(1-0) detection and the 2.6-mm continuum emission has an 
$\sim 6\sigma$ significance.
(For the continuum peak, the uncertainty is much smaller than 0.1~arcsec, as it is detected
with a much larger SNR.)
For completeness, the marginal detection from CO(1-0) emission that is kinematically associated with the opposite side of the double-horned
profile has been added to Fig.~\ref{f:cont-co} (single blue contour). This emission shows an offset of about 0.5~arcsec to the west of the 
continuum peak.
It is also noteworthy that the direction of the offset of the CO(1-0) emission peaks with respect to the 2.6-mm continuum does not 
coincide with the possible weak extension of the continuum to the south, mentioned in Section~\ref{sec:cont}.

\begin{figure}
\includegraphics[width=0.9\linewidth]{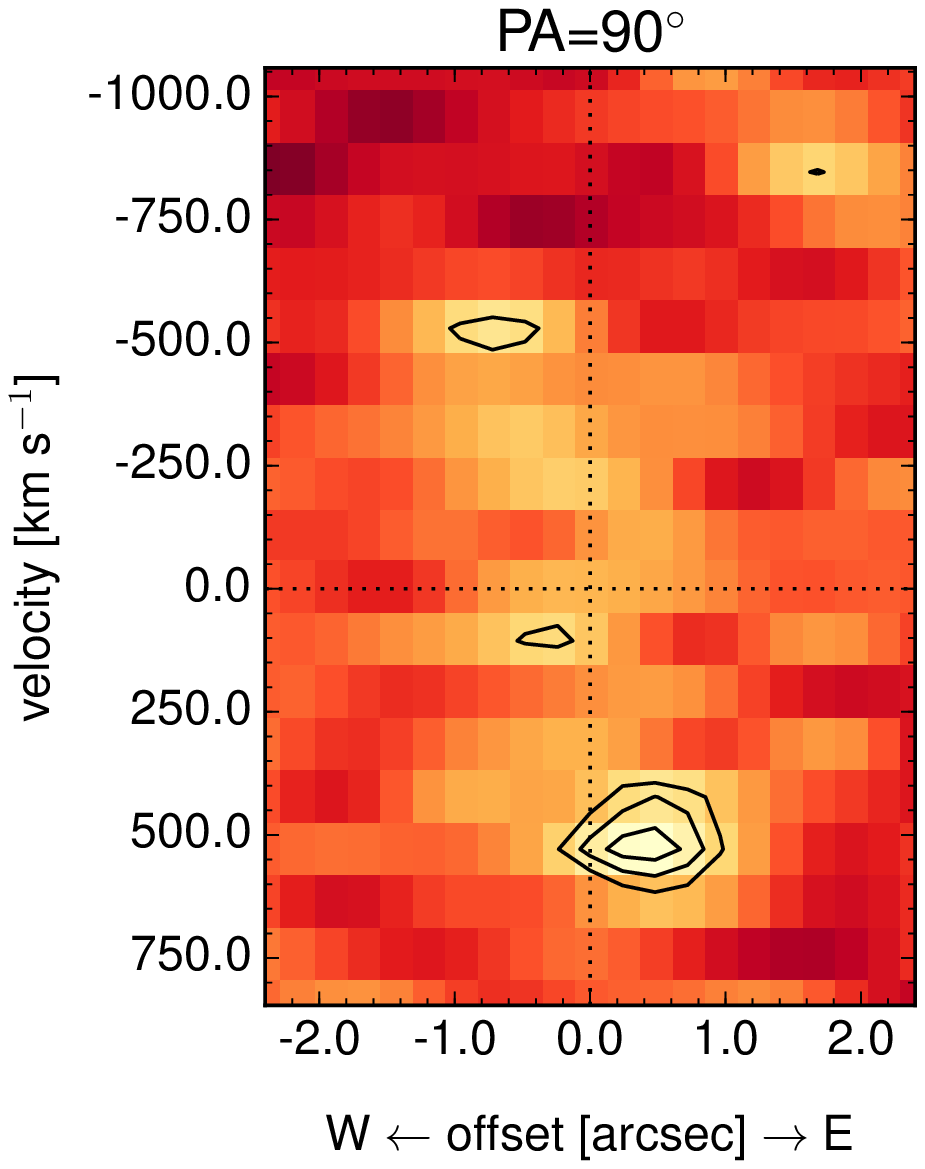}
\caption{Position-velocity diagram of the CO(1-0) emission, extracted along a 0.72-arcsec-wide slit oriented in east-west direction
and centred on the position defined by the peak of the 2.6-mm continuum. The diagram is
based on the
combined data set of all CO(1-0) observations discussed in this paper. This merged data set has a beam size of 
$1.4\times 1.1$~arcsec$^2$ with a major axis orientation along $\mathrm{PA} = 52^\circ$ and is shown at a spectral resolution of 40~MHz 
($105.8\ \mathrm{km\ s^{-1}}$) after continuum subtraction. Contours start at 4$\sigma$ and increase in steps of 1$\sigma$.\label{f:copv}}
\end{figure}

Marginal evidence of a spatial gradient across the CO(1-0) kinematic components in east-west direction is also found 
in the combined data set after merging the $1$-arcsec and $2.9$-arcsec resolution data.
The position-velocity diagram extracted from this merged data set along a slit oriented in east-west direction
is shown in Fig.~\ref{f:copv}.
Apart from the dominant $+530\ \mathrm{km\ s^{-1}}$ component, the CO(1-0) emission is characterized by a 
low signal-to-noise ratio in this spatially-resolved diagram. However, the diagram indicates that the highest velocity 
($+530\ \mathrm{km\ s^{-1}}$) component is located farthest to the east, while the lowest-velocity components around 
$-530\ \mathrm{km\ s^{-1}}$
are found farthest to the west. The evidence for a spatial gradient between the two extreme velocity components discussed in this section 
would be consistent with molecular gas clumps in 
rotation around a roughly north-south axis.

\section{Discussion}\label{sec:discussion}

\subsection{CO(1-0) kinematics and BH mass}\label{sec:COorigin}

\begin{figure}
\includegraphics[angle=-90,width=\linewidth]{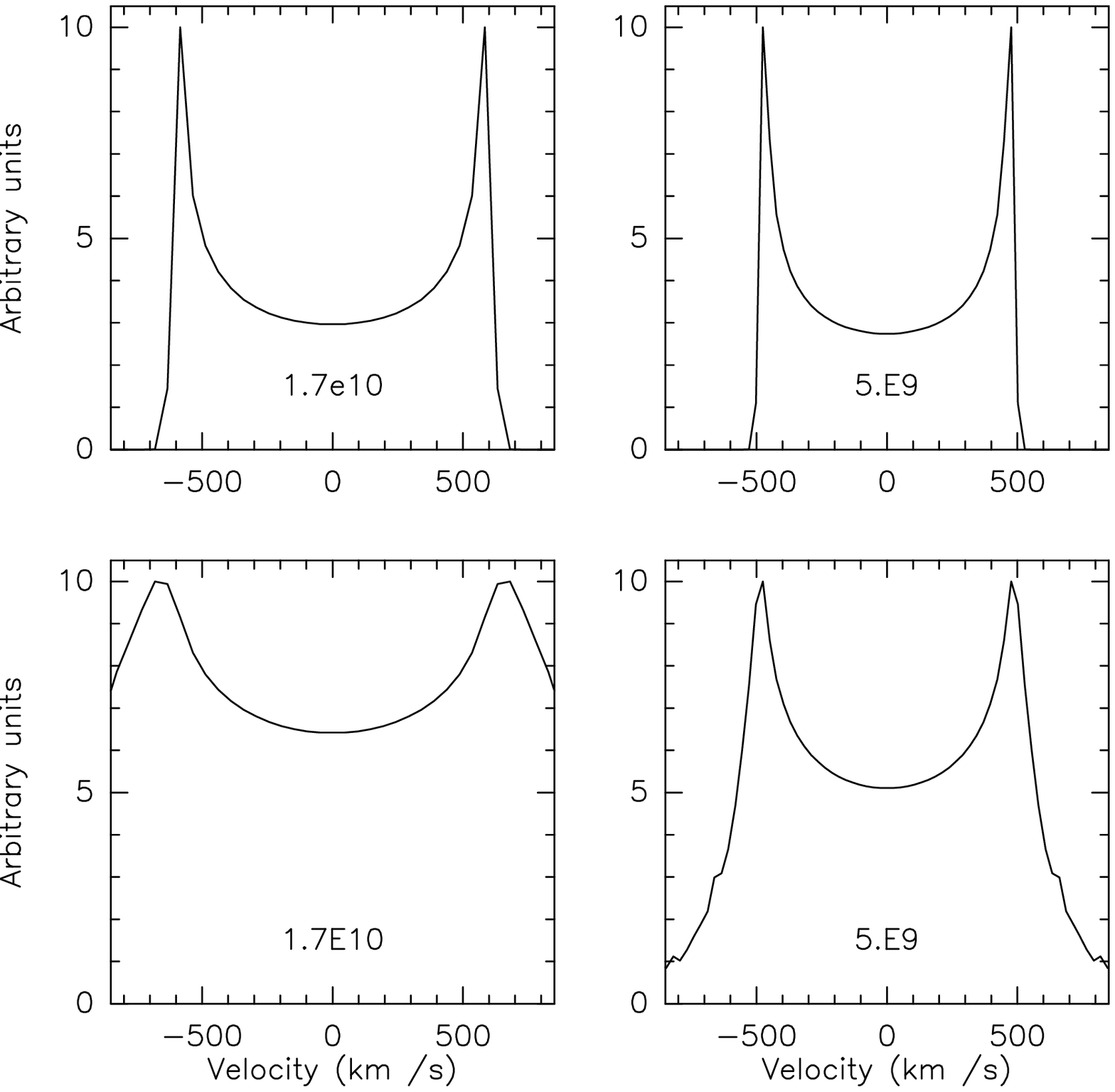}
\caption{ Molecular gas spectrum from the nuclear disc, computed
assuming the \citet{van-den-Bosch:2012aa} mass model for the stellar distribution with $M/L_V = 6.3$,
and a BH mass of $1.7\times 10^{10}\ M_\odot$ (left) and
$5\times 10^{9}\ M_\odot$ (right). The top two spectra correspond to a
gas distribution in a ring, with constant surface density between 0.28 and 0.42~kpc
(or 0.8 and 1.2~arcsec),
and the bottom two spectra to a gas distribution in an exponential disc of characteristic scale 50~pc.\label{proj-tout}} 
\end{figure}

\begin{figure}
\includegraphics[angle=-90,width=\linewidth]{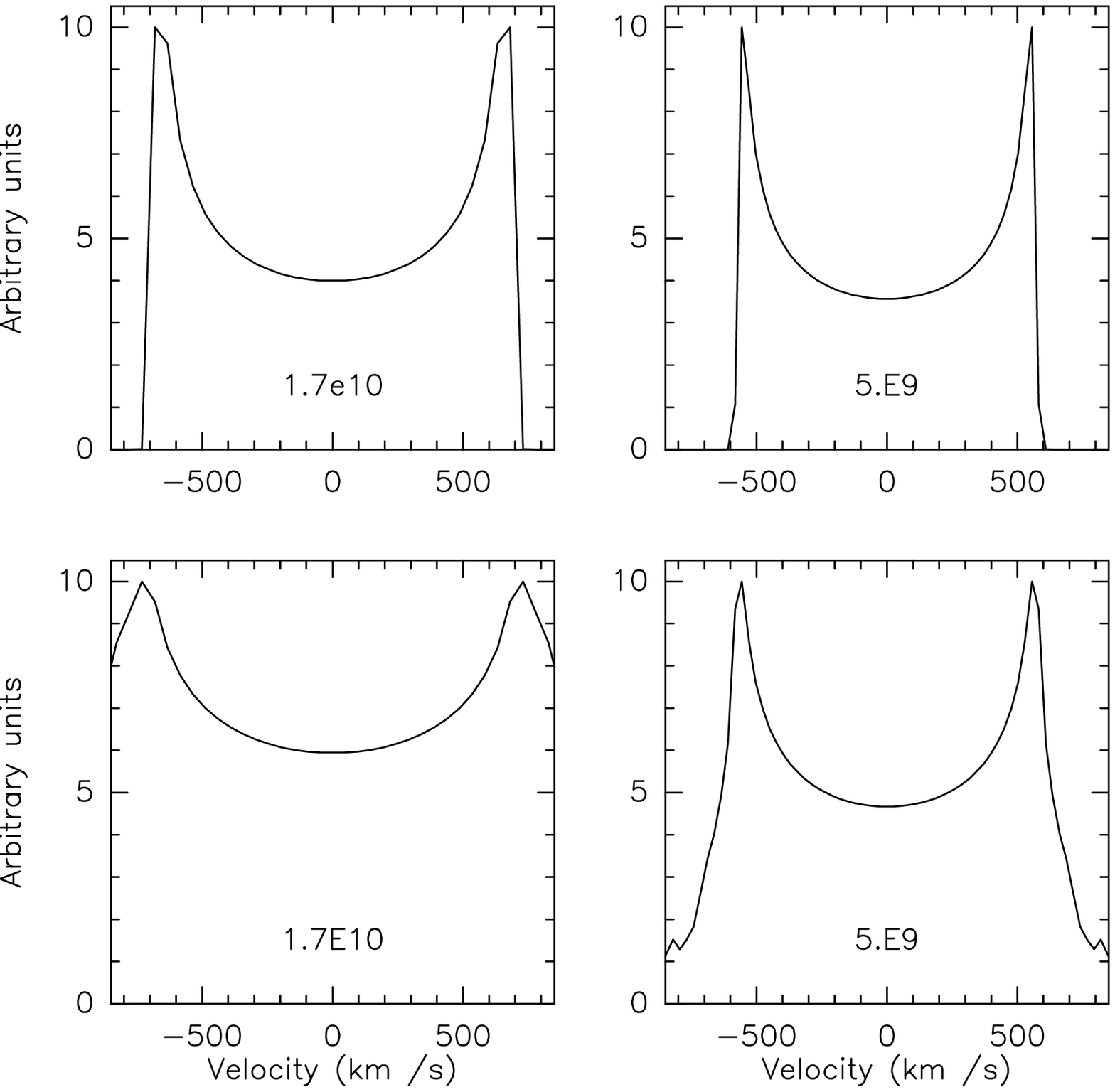}
\caption{Same as Fig.~\ref{proj-tout}, but now assuming $M/L_V = 10$.\label{proj-tout2}} 
\end{figure}

The CO(1-0) data for NGC~1277 are consistent with a scenario in which the CO(1-0)-emitting molecular gas is
distributed in a ring corresponding to the dust lane (see Fig.~\ref{fig:hst}).
The double-horned line profile shown by the CO(1-0) emission in the more compact configuration ($2.9$-arcsec resolution) 
in Fig.~\ref{f:specCDresolved} suggests that CO emission is
detected from the full extent of the dust lane.
In Figs~\ref{proj-tout} and \ref{proj-tout2}, we compare the CO(1-0) line profile to model predictions 
for different gas distributions and BH masses, based on a realistic model
of the mass distribution in the centre of NGC~1277. According to the 
data by \citet{van-den-Bosch:2012aa}, the half-light radius is 1~kpc, and the total
stellar mass is $1.2 \times 10^{11}\ M_\odot$. The distribution of light is such that 
a maximum of circular velocity of $476\ \mathrm{km\ s^{-1}}$ is reached at a radius of 0.56~kpc,
when the mass-to-light ratio is selected as $M/L_V = 6.3$. This mass distribution
corresponds also to the model of \citet{Emsellem:2013aa}, although the peak velocity is 
now $600\ \mathrm{km\ s^{-1}}$, with a higher $M/L_V = 10$.
We therefore adopted for the rotational velocity due to the stars the curve
labelled `no BH' in fig. 4 of \citet{Emsellem:2013aa},
corresponding to the best multi-Gaussian expansion fit of the light distribution. To this stellar contribution,
we added a point mass with two different values, as shown in Figs~\ref{proj-tout} and \ref{proj-tout2}.
For the molecular gas, we first assume a nuclear ring,
corresponding to the dust lane. 
The ring is modelled using a constant surface density between 0.8 and 1.2~arcsec radii,
or between 0.28 and 0.42~kpc. Since this distribution does not allow us
to sample the high velocities expected in the centre near the BH, we also consider another extreme molecular gas distribution, which is
an exponential disc with a scale of $r_0 = 50$~pc, ending also at 0.42~kpc.
For these two molecular gas distributions, we test 
two values for the BH mass, the high value of $1.7\times 10^{10}\ M_\odot$ proposed by \citet{van-den-Bosch:2012aa}
and the lower value of $5\times 10^{9}\ M_\odot$ selected by \citet{Emsellem:2013aa},
which is consistent with the most recent measurement by \citet{Walsh:2015aa}.
The inclination of the gas disc is assumed $i=75^\circ$.

The combination of these scenarios yields four possible 
spectra, which we computed according to the model already used by 
\citet{Wiklind:1997aa}.
For an assumed axisymmetric gas distribution of surface density
$n(r)$, a typical global spectrum $dN/dv (V)$ is the sum over all
radii $r$ of
\begin{equation}
\frac{dN}{dv} dV = \int{n(r) r dr d\theta} =
\int{n(r) r dr {{dV}\over{V_{\rm rot}(r) \sin{\theta} \sin{i}}}},
\end{equation}
where $V_{\rm rot}(r)$ is the rotational velocity at radius $r$,
and $i$ the inclination of the galaxy in the sky ($i=0$ is face-on).
The observed velocity in each point is
$V = V_{\rm rot} \cos{\theta} \sin{i}$, and
the global spectrum is derived by a simple
radial integration. Fig.~\ref{proj-tout} shows
the four spectra obtained when varying the BH mass
and the adopted molecular gas distribution.

The simulation clearly
shows that the most likely gas distribution is indeed the nuclear ring, corresponding to the dust lane.
For the case of an exponential gas distribution, the velocity wings of the line are boosted, which is not observed. 
The ring-like gas distribution is also supported by the preliminary results for the underlying 
spatial and kinematic structure of the CO(1-0) emission, discussed in Sect.~\ref{sec:CO} and
summarized in the form of the position-velocity diagram in Fig.~\ref{f:copv}. The emission components corresponding to the 
highest velocities around $+530$ and $-530\ \mathrm{km\ s^{-1}}$ are most enhanced and show the strongest evidence for a spatial separation.
Such a position-velocity diagram is expected for gas orbiting in an inclined ring, where the highest
line-of-sight velocities coincide with the orbital nodes of the ring, i.e. the edges of the projected lane of gas. For a smooth gas distribution,
these edges would, furthermore, be characterized by enhanced gas emission as a result of line-of-sight projection effects.
On the contrary, for gas orbiting around a central point mass in an inclined exponential disc, the highest
velocities and strongest emission would be found in the centre.

If the molecular gas is distributed in the dust lane, the rotational velocities probe the total enclosed mass.
The two horns of the observed CO(1-0) profile in Fig.~\ref{f:specCDresolved} are roughly located at
$V = +530$ and $-530\ \mathrm{km\ s^{-1}}$ with respect to the systemic velocity of NGC~1277.
For an inclination of $i=75^\circ$, this corresponds to a rotational velocity of 
$V_\mathrm{rot}=V/\cos(90^\circ-i)\approx 550\ \mathrm{km\ s^{-1}}$.
The enclosed mass 
is given by $M_\mathrm{enc}= (V_\mathrm{rot}^2 r)/G$, where $r$ is the distance of the gas from the BH and $G$ is the
gravitational constant.
Using a distance of $r\sim 320$~pc (0.9~arcsec), corresponding to the radius of the dust lane, the resulting total enclosed 
mass is $M_\mathrm{enc}\sim 2\times10^{10}\ M_\odot$. The models in Figs~\ref{proj-tout} and \ref{proj-tout2}
provide a more detailed analysis by distinguishing the stellar and BH mass contributions.
Compared to the line profile in Fig.~\ref{f:specCDresolved}, the best model for $M/L_V = 6.3$ 
(Fig.~\ref{proj-tout}) appears to be the top left one,
supporting the high mass for the BH. 
For $M/L_V = 10$ (Fig.~\ref{proj-tout2}), the width of the observed CO(1-0) spectrum can also be reproduced by the 
$5\times 10^{9}\ M_\odot$ BH (top right panel).
We conclude that the observed CO(1-0) line profile is consistent with a $1.7\times 10^{10}\ M_\odot$ BH, while a lower mass of
$5\times 10^{9}\ M_\odot$ is likewise possible, if the underlying mass distribution is characterized by $M/L_V = 10$. 

\begin{figure}
\centering
\includegraphics[width=\linewidth]{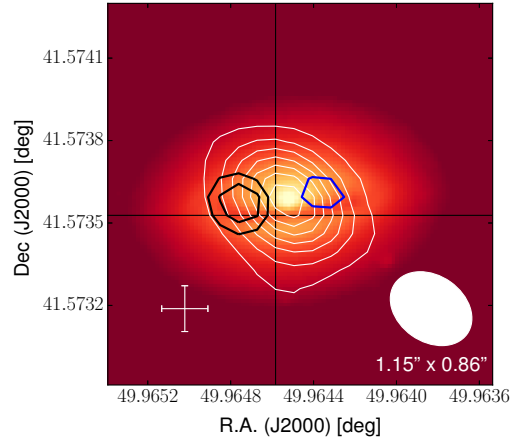}
\caption{Overlay of the CO(1-0) line and 2.6-mm continuum emission found at 1-arcsec spatial resolution and the {\it HST} image. 
For this overlay, the astrometry of the {\it HST} image
has been adjusted by hand in order to match the scenario, in which the 2.6-mm continuum peak is associated with the optical galaxy nucleus.
The {\it HST} image is shown in colour scale. The 2.6-mm continuum contours (white) and the 
contours for the two CO(1-0) peaks (black and blue) are the same as in Fig.~\ref{f:cont-co}.
The size and orientation of the beam corresponding to the continuum and CO(1-0) map is indicated in the lower right corner.
The error bar in the lower left corner shows the estimated uncertainty of $\pm 0.3$~arcsec 
in the absolute astrometry of the {\it HST} image. \label{f:HSTcontco}}
\end{figure}

In the data at $1$-arcsec spatial resolution, the only clear CO(1-0) detection is 
associated with the pronounced kinematic component around $+530\ \mathrm{km\ s^{-1}}$.
This emission peak is found to be offset by $\sim 0.6$~arcsec largely to the east
of the 2.6-mm continuum peak. 
It could originate from a molecular gas clump close to the eastern orbital node of the dust lane.
As discussed in Sect.~\ref{sec:CO}, we also find marginal evidence of a CO(1-0) emission
peak on the opposite side of the 2.6-mm continuum peak associated with similar absolute, but blueshifted, velocities. This emission  
may originate from a region close to the western orbital node of the dust lane.
The H$_2$ masses derived for the eastern and western emission peaks are $\sim 2\times 10^7$ and $\sim 1\times 10^7\ M_\odot$, respectively
(see Table~\ref{tab:co}). These values are at the upper end of the gas masses found for the most massive giant molecular clouds in the Milky Way \citep[e.g.][]{Murray:2011aa}.
In addition to the molecular cloud scenario, it is also possible that the peaks
are enhanced by line-of-sight crowding effects, which would be expected around the orbital nodes of
an inclined ring model.
The above interpretation, in which the CO(1-0) peaks are associated with the orbital nodes of the dust lane, 
is highlighted in comparison to the {\it HST} image in Fig.~\ref{f:HSTcontco}.
While a blind comparison between the millimetre data and the {\it HST} image is subject to 
the astrometric uncertainty of $\sim0.3$~arcsec in the {\it HST} image, the astrometry in this figure
has been adjusted by hand in order to align the 2.6-mm continuum peak with the optical galaxy nucleus.

\subsection{The molecular gas as a reservoir for star formation}\label{sec:starformation}
The CO(1-0) emission discussed here indicates a reservoir of molecular gas in the central region of NGC~1277, which can potentially fuel star formation.
The H$_2$ mass of $1.5\times 10^8\ M_\odot$ (Table~\ref{tab:co}) corresponding to the double-horned profile seen at 2.9-arcsec resolution is in good agreement 
with the typical H$_2$ masses found
for early-type galaxies from ATLAS$^\mathrm{3D}$ \citep{Young:2011aa}, but lower than the 
integrated H$_2$ masses found for dust-lane early-type galaxies \citep[$4\times 10^8$ to $2\times 10^{10}\ M_\odot$, ][]{Davis:2015aa}.
A rough estimate of the mean surface density of the $1.5\times 10^8\ M_\odot$
of molecular gas in NGC~1277 results in about $500\ M_\odot\ \mathrm{pc^{-2}}$.
According to the Kennicutt-Schmidt relation \citep{Schmidt:1959aa, Kennicutt:1998aa}, this could imply a star formation rate density of 
$\sim 0.5\ M_\odot\ \mathrm{yr^{-1}\ kpc^{-2}}$ and an integrated star formation rate of the order of $0.1\ M_\odot\ \mathrm{yr^{-1}}$.
For this estimate, the size of the molecular gas reservoir and star formation region is assumed to be the ring confined to radii between
0.8 arcsec and 1.2 arcsec, consistent with the model presented in Sect.~\ref{sec:COorigin}. However, the exact dimensions
of the ring are unknown, so that this estimate involves large uncertainties. In addition, it also has to be taken into account that
the mean gas surface density may not be a good representation of the actual surface density, since the gas distribution could be clumpy. 
Further insights into the clumpiness of the gas distribution could be obtained from high resolution observations.
However, based on the current sensitivity limit of the data obtained at 1-arcsec resolution, it remains unclear whether the observed CO(1-0) emission peaks 
rather originate from giant molecular clouds or from 
line-of-sight projection effects.

UV data from the {\it GALEX} \citep{Martin:2005aa} archive independently suggest a low level of active
star formation in NGC~1277 with a
global star formation rate of the order of $0.1\ M_\odot\ \mathrm{yr^{-1}}$. 
The far-UV (FUV) flux reported for NGC~1277 in the {\it GALEX} archive is
$(20.24\pm0.52)\ \mu\mathrm{Jy}$, or $(67.2\pm1.7)\ \mu\mathrm{Jy}$ after correction for foreground extinction 
using $E(B-V)=0.165$ from \citet{Schlegel:1998aa} and $A_\mathrm{FUV} = 7.9 E(B-V)$ \citep{Gil-de-Paz:2007aa}. This flux appears to be associated
with a slightly extended detection in the far- and near-UV images. 
The corresponding luminosity is $L_\mathrm{FUV}=8.9\times 10^{41}\ \mathrm{erg\ s^{-1}}$.
Together with the total infrared luminosity of NGC~1277 of $L_\mathrm{IR} (8-1000\ \mu \mathrm{m}) = 2.3 \times10^9\ L_\odot = 8.8\times 10^{42}\ \mathrm{erg\ s^{-1}}$, which
we will derive in Sect.~\ref{sec:continuumorigin}, the FUV detection indicates a dust-corrected star formation rate of about $0.2\ M_\odot\ \mathrm{yr^{-1}}$, using the
calibration 
\begin{equation}
\mathrm{SFR}=4.6\times 10^{-44}[L_\mathrm{FUV}+0.46L_\mathrm{IR}(8-1000\ \mu \mathrm{m})]
\end{equation} 
\citep{Hao:2011aa,Calzetti:2013aa}. Note that this calibration is based on a Kroupa initial mass function, while a Salpeter initial mass function would result
in a 55 per cent larger star formation rate \citep[e.g.][]{Smith:2015aa}. 
The actual star formation rate may be lower than the value derived above, since the FUV flux could include contributions from
an old stellar population \citep[cf. ][]{Burstein:1988aa} or an AGN.
In Sections~\ref{sec:continuumorigin} and \ref{sec:accretion}, we discuss indications for an AGN in NGC~1277. 
However, the optical spectra of NGC~1277 presented by
\citet{Trujillo:2014aa} and \citet{Martin-Navarro:2015aa} do not 
show any prominent AGN emission lines, so that an AGN, if present, is likely to be weak.

The FUV emission in star forming regions is dominated by 
young ($\lesssim 100$~Myr) massive stars. Therefore, the star formation history derived from optical spectra may not be sufficiently sensitive to a low-level
of recent star formation with a rate of a few $0.1\ M_\odot\ \mathrm{yr^{-1}}$ in a stellar population otherwise dominated
by old stars. The stellar population properties based on optical spectra for NGC~1277 indicate a 
uniformly old $\gtrsim 10$~Gyr stellar population
\citep{van-den-Bosch:2012aa,Trujillo:2014aa,Martin-Navarro:2015aa}. Furthermore, 
the optical spectra presented by \citet{Martin-Navarro:2015aa} in their Fig.~2 do not provide any strong evidence of H$\alpha$ emission as an indicator of
active star formation. By visual inspection, the figure only shows a possible emission peak close to the H$\alpha$ wavelength in the nuclear averaged spectrum, which
could be associated with weak emission from ionized gas.

Since the above estimates for the star formation rate in NGC~1277 involve large uncertainties, 
we also have to consider the possibility that star formation is suppressed despite the
presence of a $1.5\times 10^8\ M_\odot$ reservoir of molecular gas. Shear can have a disruptive effect on star formation by altering the collapse and fragmentation 
of molecular clouds \citep[e.g.][]{Hocuk:2011aa, Dobbs:2013aa}. High shear has been discussed as a possible explanation for the
lower star formation efficiency with respect to the Kennicutt-Schmidt relation, displayed by early-type galaxies from the ATLAS$^{\mathrm{3D}}$ sample
\citep{Davis:2014ab}. In the steeply rising gravitational potential towards the centre of NGC~1277, shear can potentially play a role in
preventing efficient star formation in molecular clouds in the region of the dust lane.

Active star formation would tend to lower the $V$-band stellar mass-to-light ratio compared to a uniformly old stellar population. 
As discussed in Sect.~\ref{sec:COorigin}, a tendency towards a lower $M/L_V$ would be in favour of a larger BH mass, 
because the required velocity width of the CO(1-0) line profile can only be reproduced when compensating the smaller stellar mass contribution with a
larger central BH mass (see Figs~\ref{proj-tout} and \ref{proj-tout2}).
However, recent observational data are rather in support of a high mean stellar mass-to-light ratio in the central region.
Based on optical long-slit spectra obtained in 0.8-arcsec seeing, 
\citet{Martin-Navarro:2015aa} suggest that the stellar mass-to-light ratio in NGC~1277 increases towards the nucleus.
These data are limited in spatial resolution and lack detailed information
on the run of the stellar mass-to-light ratio inside the region encompassed by the dust lane.
The innermost off-nuclear radial bin in their data 
roughly corresponds to the radius of the dust lane.
A high stellar $V$-band mass-to-light ratio of $M/L_V=(9.3\pm1.6)$ for the central 1.3 arcsec in NGC~1277 has also been 
reported by \citet{Walsh:2015aa}, based on high-spatial-resolution near-infrared integral field spectroscopy.
 This value is derived from stellar-dynamical models assuming a constant mass-to-light ratio, so that possible radial variations in 
 the mass-to-light ratio are not traced.

\subsection{Origin of the 2.6-mm continuum emission}\label{sec:continuumorigin}

\begin{figure*}
\centering
\includegraphics[width=0.7\linewidth]{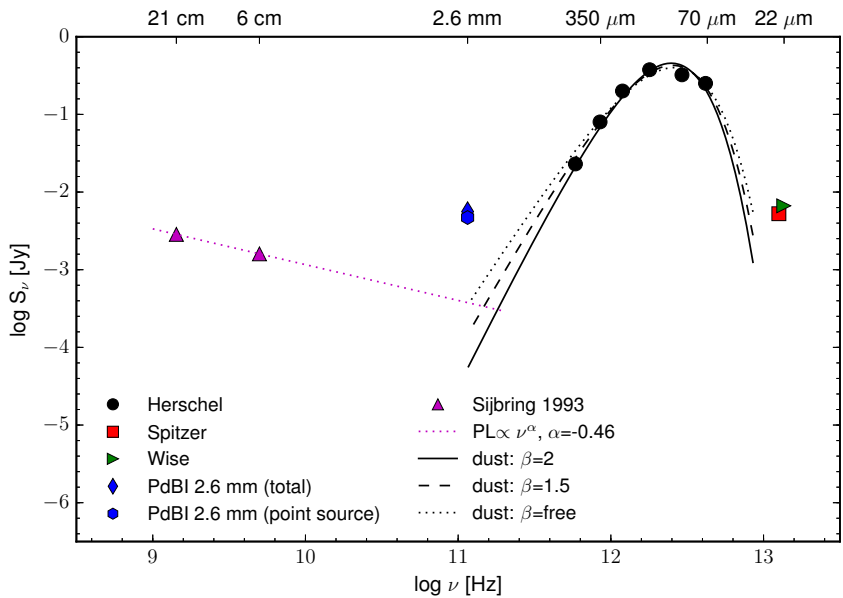}
\caption{Infrared-to-radio spectral energy distribution of NGC~1277, together with an example model fit based on the far-infrared
and radio data (see text for details). The infrared measurements are compiled based on 
data from {\it Herschel}, {\it Spitzer}, and {\it Wise}, as described in the text. For the 2.6-mm continuum 
measured in this work the figure shows both, the total aperture-integrated flux as well as the point source contribution discussed
in Sect.~\ref{sec:cont}. The 1.4 and
5~GHz fluxes 
are taken from \citet{Sijbring:1993}. The solid, dashed, and dotted black lines show the fits for the dust component, 
assuming $\beta=2$, $\beta=1.5$, and a free $\beta$, as described in the text, extrapolated to 2.6 mm. The dotted (magenta) line connecting the 1.4 and 5 GHz data points shows the results from fitting
a power-law $S_\nu \propto \nu^\alpha$
to the 1.4 and
5~GHz fluxes. 
\label{f:sed}}
\end{figure*}

The 2.6-mm continuum flux of NGC~1277 of $(5.6 \pm 0.2)$~mJy 
could be associated with star formation or with an AGN-related radio jet. In order to discuss the origin of this continuum emission, 
we show the 2.6-mm continuum flux together with the infrared-to-radio spectral energy distribution of NGC~1277 in Fig.~\ref{f:sed}. 
The analysis of the spectral energy distribution is likely to involve uncertainties given that the composite is
based on an inhomogeneous data set. Furthermore, the significance of the models discussed in the following is limited by the small number of published data at
radio frequencies. To our knowledge, the only published data in the radio domain are the 1.4 and 5~GHz detections of 2.85 and 
1.6~mJy, respectively, reported by \citet{Sijbring:1993}, who found the radio source to be unresolved in the corresponding $13\times 19$~arcsec$^2$
and $3.5\times 5.2$~arcsec$^2$ beams. 

Fig.~\ref{f:sed} shows example models for the available data, based on the far-infrared and radio continuum, which under-predict 
the flux at 2.6 mm by about one order of magnitude.
The infrared data include data from {\it Herschel}, {\it Spitzer}, and {\it Wise}. The {\it Herschel} photometry measurement was performed using the
Herschel Interactive Processing Environment (HIPE) 13.0.0.
The SPIRE fluxes are from {\sc SUSSEXtractor}, while the results
from the {\sc Timeline Fitter} agree within a few per cent.
The PACS fluxes are derived from aperture photometry with
aperture correction included.
We used the python code {\sc mbb\_emcee} to fit modified blackbodies
to photometry data using an affine invariant Markov chain Monte Carlo (MCMC) method, with the
{\it Herschel} passband response folded
\citep{Dowell:2014aa}.
Assuming that all dust grains share a single temperature $T_d$, that
the dust distribution is optically thin, and neglecting any power-law component
towards shorter wavelengths, the fit results in a temperature of 
$T_d/(1+z)= (24.0 \pm 0.8)$~K, 
a luminosity $L_\mathrm{IR} (8-1000\ \mu \mathrm{m}) = (2.3\pm0.2) \times10^9\ L_\odot$, 
and a dust mass of $M_d=(2.6\pm 0.5) \times 10^6\ M_\odot$ for $\beta =2$. 
For $\beta =1.5$, the corresponding values are
$T_d/(1+z)= (27 \pm 1)$~K, 
$L_\mathrm{IR} (8-1000\ \mu\mathrm{m}) = (2.4\pm0.2) \times10^9\ L_\odot$, 
and $M_d=(1.3\pm 0.2) \times 10^6\ M_\odot$. 
We also performed a fit using a free $\beta$ parameter, which results in 
$T_d/(1+z)= (32 \pm 5)$~K, 
$L_\mathrm{IR} (8-1000\ \mu\mathrm{m}) = (2.4\pm0.2) \times10^9\ L_\odot$, 
$M_d=(9\pm 5) \times 10^5\ M_\odot$, and $\beta=1.1\pm0.4$.
The dust temperatures are similar to the ones found for the cold-to-warm dust
component in galaxies. The cold component ($\sim15$ - 25~K) has been interpreted as diffuse dust heated by the general interstellar radiation field from young and/or old
stellar populations, while the warmer component ($\sim30$ - 40~K) is assumed to be associated with dust in star forming regions heated by young O and B stars
\cite[e.g.][]{Cox:1986aa}. The uniformly old stellar population of NGC~1277 identified via optical spectroscopy 
\citep{van-den-Bosch:2012aa, Trujillo:2014aa,Martin-Navarro:2015aa} is likely to provide
heating of diffuse dust by the general interstellar radiation field. As discussed in Sect.~\ref{sec:starformation},
NGC~1277 may also host a low level of active star formation, which could contribute a warmer dust component.
However, it is evident that the cold-to-warm dust component derived from the model in Fig.~\ref{f:sed} 
can clearly not account for the continuum emission found at 2.6~mm.

The radio fluxes at 1.4 and 5~GHz have been fitted with a single power-law of the form $S_\nu \propto \nu^\alpha$, resulting in
a power-law index of $\alpha = -0.46$. This index is in
agreement with a synchrotron spectrum. 
Given that the stellar population of NGC~1277 is dominated by old stars
\citep{van-den-Bosch:2012aa, Trujillo:2014aa,Martin-Navarro:2015aa} and that recent star formation
is only present at a low level at most (Sect.~\ref{sec:starformation}), it is likely that the 1.4 and 5~GHz data trace non-thermal emission from an extended radio jet. 
This conclusion is also supported by the fact that the given radio and far-infrared fluxes
indicate a mild radio excess with respect to the radio-far-infrared relation
for star forming galaxies. Based on the definition
\begin{equation}
q = \log\left[\frac{ S_\mathrm{FIR}}{3.75\times 10^{12}\ \mathrm{W\ m^{-2}}} \right]- \log\left[ \frac{S_{1.4\ \mathrm{GHz}}}{\mathrm{W\ m^{-2}\ Hz^{-1}}} \right]
\end{equation}
with $S_\mathrm{FIR} = 1.26\times 10^{-14} [2.58\ S_\mathrm{60\ \mu m} + S_\mathrm{100\ \mu m}]\ \mathrm{W\ m^{-2}}$ from \citet{Yun:2001aa},
the 70 and 100 $\mathrm{\mu m}$ fluxes for NGC~1277 from {\it Herschel} and the 1.4~GHz flux from \citet{Sijbring:1993} suggest $q\sim 2.0$.
A comparison with fig. 6 in \citet{Yun:2001aa} shows that this $q$-value indicates some radio excess compared to the distribution of $q$-values 
for far-infrared selected galaxies and, therefore, a possible contribution from an AGN.
In other words, using the 1.4-GHz flux as an indicator of the star formation rate based on the radio-far-infrared correlation
\citep[e.g. ][]{Yun:2001aa, Murphy:2011aa}, results
 in a recent star formation rate that is about one order of magnitude larger 
 than the rates discussed in Sect.~\ref{sec:starformation}.

The excess of the 2.6-mm continuum in Fig.~\ref{f:sed} (both, when considering the total flux or only the point source contribution) 
with respect to the synchrotron-like radio continuum derived from the 1.4 and 5~GHz fluxes of NGC~1277
is about one order 
of magnitude. 
The 2.6-mm flux indicates an inverted spectral energy distribution from the
radio to the millimetre regime, which cannot be fully explained by $S_\nu \propto \nu^{-0.1}$
free-free emission from \ion{H}{ii} regions \citep[cf. ][]{Murphy:2011aa} and rather suggests AGN emission.
Some of this excess at 2.6 mm could be caused by variability or by extended non-thermal emission probed more 
prominently by the larger beam size corresponding to the
1.4 and 5~GHz data. Similarly inverted spectra
have been found for a number of low-luminosity AGN and 
elliptical galaxies \citep{Doi:2005aa, Doi:2011aa}. 
These authors discuss the possibility that the strong flux at millimetre wavelengths is a signature of the accretion disc in 
an advection-dominated accretion flow \citep{Ichimaru:1977aa}.
More data at radio and millimetre wavelengths will be required for a more stringent interpretation of the
radio-to-millimetre spectral energy distribution of NGC~1277.

The presence of a weak AGN in NGC~1277 would be consistent with X-ray data. 
\citet{Fabian:2013aa} find evidence for an unresolved power-law source with a 0.5--7~keV luminosity of  
$1.3\times 10^{40}\ \mathrm{erg\ s^{-1}}$. A rough estimate indicates that an AGN producing the X-ray and the non-thermal radio emission would be in agreement with
the known radio-X-ray correlations for AGN. The 6-cm (5~GHz) flux of $S_\mathrm{6\ cm} = 1.6$~mJy from \citet{Sijbring:1993}
corresponds to a luminosity of $L_\mathrm{6\ cm} = 5.5\times 10^{37}\ \mathrm{erg\ s^{-1}}$.
Using the 0.5--7~keV as an approximation for the 2--10~keV luminosity, the radio and X-ray fluxes are consistent with the 
radio-X-ray correlation, which is, e.g., shown by the corresponding projection of the Fundamental Plane of BH activity in \citet[][left-hand panel of
their fig. 3]{Merloni:2003aa}.

\subsection{Implications for gas accretion on to the BH}\label{sec:accretion}

If NGC~1277 hosts a BH with a mass of $\sim 5\times 10^{9}$ to $10^{10}\ M_\odot$,
this BH is overmassive compared to the value expected from the Fundamental Plane of BH activity \citep{Merloni:2003aa}, assuming
that the radio continuum as well as the X-ray power-law source are indeed associated with an AGN.
According to the Fundamental Plane relation from \citet{Gultekin:2009ab}, 
the 6-cm radio luminosity of $5.5\times 10^{37}\ \mathrm{erg\ s^{-1}}$
(see Section~\ref{sec:continuumorigin}) combined with an approximate 2--10~keV luminosity of  
$1.3\times 10^{40}\ \mathrm{erg\ s^{-1}}$ \citep[][and Section~\ref{sec:continuumorigin}]{Fabian:2013aa} suggests
a BH mass of about $1\times 10^8\ M_\odot$, which is two orders of magnitude smaller than the BH mass of $1.3 \times 10^{10}$ to $1.7\times 10^{10}\ M_\odot$ 
suggested by \citet{van-den-Bosch:2012aa}
and \citet{Yildirim:2015aa} and still significantly smaller than the revised BH mass measurement of 
$(4.9\pm 1.6)\times 10^9\ M_\odot$ from \citet{Walsh:2015aa}.

There is evidence that the gas accretion on to the BH in NGC~1277 deviates from simple Bondi accretion.
Based on the thermal gas detected within the Bondi radius, \citet{Fabian:2013aa} show that the accretion luminosity expected from 
radiatively efficient Bondi accretion 
is 5--6 orders of magnitude more luminous than the observed X-ray power-law component.
As pointed out in Sect.~\ref{sec:continuumorigin}, a similar order-of-magnitude difference is suggested by
the 6-cm flux from \citet{Sijbring:1993}, which is consistent with
the radio-X-ray correlation as one projection of the Fundamental Plane of BH activity (though for a much smaller BH).
Why the observed X-ray luminosity of the central point source in NGC~1277 is much smaller than the one expected for 
radiatively efficient Bondi accretion, remains
unclear.
\citet{Fabian:2013aa} suggest the possibility of a very radiatively-inefficient accretion process. Alternatively, gas angular momentum 
may be responsible for lowering the accretion rate \citep{Li:2013aa}.

The dust lane is likely to play an important role in impeding cold gas accretion. This has already been pointed out by \citet{Fabian:2013aa}, considering the
high velocities in the dust lane that would be expected from the large BH mass.
Our data support this picture, since the CO(1-0) detection is consistent with emission from a ring and shows the predicted high rotational 
velocities.
While the gas in the dust lane may eventually loose angular momentum due to viscous processes and begin to spiral inward, the current lack of cold gas in the
nuclear region may be the reason for the very low current BH accretion rate.

\section{Summary and conclusions}\label{sec:summary}

We have reported a detection of CO(1-0) emission from NGC~1277, probed at $\sim 1$ and $\sim 2.9$~arcsec
spatial resolution, using the IRAM PdBI. The data indicate that the molecular gas is distributed in the ring of the dust lane 
encompassing the nucleus of NGC~1277 at a distance of 0.9~arcsec. Furthermore, the molecular gas shows high rotational velocities
of $\sim 550\ \mathrm{km\ s^{-1}}$.
This is suggested consistently by the low- and high-resolution 
data. The low-resolution data reveal a very broad, double-horned line covering a symmetric velocity range around
the systemic velocity of NGC~1277, which appears marginally resolved in terms of kinematics in east-west direction.
In the high-resolution data, the pronounced red horn of the double-horned profile, centred at 
$\sim +530\ \mathrm{km\ s^{-1}}$, is detected with an offset of 
$\sim 0.6$~arcsec
to the east of the 2.6-mm continuum peak, while a marginal detection of kinematic components associated with 
the blue horn are found to the west of the continuum peak.
The high rotational velocities of the molecular gas provide independent evidence for a central mass concentration
of $\sim 2 \times10^{10}\ M_\odot$ inside the radius of the dust lane. This enclosed mass is large enough to potentially 
host an ultramassive BH as massive as reported by \citet{van-den-Bosch:2012aa} and \citet{Yildirim:2015aa} based on stellar kinematics.
Compared to models with realistic mass distributions for $M/L_V = 6.3$, the spatially-unresolved CO(1-0) line profile is
consistent with the profile expected for the presence of a 
central $1.7 \times10^{10}\ M_\odot$ BH. However, a lower-mass BH of $5 \times10^{9}\ M_\odot$, consistent with the revised
measurement by \citet{Walsh:2015aa}, is also possible, if the
stellar mass-to-light ratio is assumed to be $M/L_V = 10$.

Future observations are required in order to further resolve this ambiguity in the gas-kinematic measurement.
Tighter gas-kinematic constraints on the BH mass can be expected from sensitive observations
at high angular resolution, by probing gas clouds as close as possible
to the BH (i.e. the gas with the highest rotational velocities). If the molecular gas is indeed 
largely confined to a ring, such observations will ultimately be limited by the lack of gas in the vicinity of the BH.
However, depending on the radial extent of the ring, gas kinematics from the innermost parts of the ring may already 
be sufficiently sensitive to the central gravitational potential to provide a clearer distinction between the BH mass and extended stellar mass contributions.
A better knowledge of the stellar mass-to-light ratio in the circum-nuclear region 
could similarly improve the gas-kinematic
BH mass constraints. 
Stellar absorption line studies for NGC~1277 have so far been limited in spatial resolution \citep[see, e.g., the discussion by][]{Yildirim:2015aa}, so that the
detailed run of stellar mass-to-light ratio with radius inside the region encompassed by the dust lane remains uncertain.
Recent results by \citet{Walsh:2015aa} indicate a high stellar mass-to-light ratio of 
$M/L_V = 9.3\pm 1.6$ for the central region of NGC~1277.
This measurement is based on stellar-dynamical
models assuming a constant mass-to-light ratio out to a radius of 1.3 arcsec ($\sim 460$~pc), which covers the full extent
of the dust lane. This value would be in favour of an $\sim 5 \times10^{9}\ M_\odot$ gas-kinematic BH mass,
in agreement with the stellar-dynamical mass derived by \citet{Walsh:2015aa}.

The strong underlying 2.6-mm continuum emission of NGC~1277 cannot be explained by the cold dust component seen
in the far-infrared and appears to be in excess of the radio-to-millimetre spectral energy distribution expected for free-free emission
from \ion{H}{ii} regions.
Based on literature X-ray and radio data, we suggest that the 2.6-mm continuum is caused by a weak AGN. The continuum is extended and shows marginal evidence
of an extension to the south that could be related to a radio-jet.

If the BH in NGC~1277 is as massive as $\sim 5\times 10^9$ to $\sim1.7 \times10^{10}\ M_\odot$, it appears to be an overmassive 
outlier with respect to the Fundamental Plane of BH activity as well as the $M_\mathrm{BH}$-$L_{\mathrm{sph,}K}$-relation -- although the
latter may strongly depend on the details of the bulge-disc decomposition \citep{Savorgnan:2015ab}. 
It is likely that the peculiar properties of NGC~1277 are a result of its location in the environment of the Perseus Cluster.
It has been suggested that the
 formation of ultramassive BHs may have proceeded rapidly in the early Universe, well before $z\sim 2$--3 \citep{Dubois:2012aa, Fabian:2013aa}. Early BH growth
 in NGC~1277
is supported by the fact that NGC~1277 seems to currently lack any significant BH accretion.
The offset from the $M_\mathrm{BH}$-$L_{\mathrm{sph,}K}$-relation indicates that NGC~1277 has either lost a large part of its stellar mass
or has not been able to acquire additional stellar mass and gas after an initial phase of star formation. Furthermore, the predominantly old stellar population, the
currently small BH accretion rate, and the offset from the Fundamental Plane of BH activity suggest that NGC~1277 has been devoid of gas during its recent evolution.
A stripping event in the cluster environment could have removed large parts of the gas and stellar mass, 
although this scenario has so far been regarded less likely
\citep[e.g.][]{van-den-Bosch:2012aa}. 
If NGC~1277 has been moved off the $M_\mathrm{BH}$-$L_{\mathrm{sph,}K}$-relation by tidal stripping, this implies that
galaxies with ultramassive BHs have been close to, or in agreement with, this relation by $z\sim$ 2--3 and that 
ultramassive BHs have primarily evolved in lockstep with their host galaxies. 
Alternatively, as suggested by \citet{Fabian:2013aa},
NGC~1277 may have undergone quenching at high redshift and may have been prevented
from acquiring further gas and stellar mass 
by its off-centre location in the Perseus Cluster, in contrast to central cluster galaxies which may have grown significantly through
mergers. 
In this case, NGC~1277 would have neither experienced galaxy nor BH growth in its more recent evolution and may be a `relic' galaxy \citep{Trujillo:2014aa}, in terms of both, its stellar population and BH. This scenario has recently been explored in more detail by 
\citet{Ferre-Mateu:2015aa}. As these authors discuss, this scenario 
implies that the growth of ultramassive BHs may have preceded the growth of their host galaxies, so that
all galaxies with ultramassive BHs may have been outliers from the BH-bulge luminosity (mass) relation by $z\sim$ 2--3, in a similar 
manner as still observed for NGC~1277 today. 
It is noteworthy
that the BH in the central galaxy of the Perseus Cluster, NGC~1275, is significantly less massive than the one 
suggested for NGC~1277
\citep{Scharwachter:2013aa}. As a possible explanation, \citet{Shields:2013ab} proposed that NGC~1277 may have captured a 
massive BH that had originally grown in NGC~1275. 

Regardless of which process may have led to this particular BH distribution in NGC~1275 and NGC~1277,
it is evident that probing the most massive BHs in clusters will provide new insights into cluster formation and evolution. 
Given the very small number of confirmed ultramassive BHs to-date, future observations are required
in order to better understand the growth of the most massive BHs and their host galaxies as well as the role of the cluster environment.

\section*{Acknowledgements}

We thank the anonymous referee for a constructive report and interesting suggestions which have led to Sect.~\ref{sec:starformation}.
This work is based on observations carried out under projects number X094 and W14DB with the IRAM Plateau de Bure Interferometer. IRAM is supported by INSU/CNRS (France), MPG (Germany) and IGN (Spain). We are grateful to the IRAM staff, especially Sabine K\"onig, for supporting the observations.
JS and FC acknowledge the European Research Council
for the Advanced Grant Program Num 267399-Momentum.
This research has made use of the NASA/IPAC Extragalactic Database (NED) which is operated by the Jet Propulsion Laboratory, California Institute of Technology, under contract with the National Aeronautics and Space Administration. 
Some of the data presented in this paper were obtained from the Mikulski Archive for Space Telescopes (MAST). STScI is operated by the Association of Universities for Research in Astronomy, Inc., under NASA contract NAS5-26555. Support for MAST for non-HST data is provided by the NASA Office of Space Science via grant NNX09AF08G and by other grants and contracts.

\section*{Note added in proof}
Based on a new bulge/disc decomposition and new stellar-kinematic data, \citet{Graham:2016aa} propose that NGC 1277 might be brought back on the BH-to-host spheroid mass relation, albeit with a high $M/L_V= 12.3$.

\bibliographystyle{mnras}
\bibliography{ngc1277bib}

%%%%%%%%%%%%%%%%%%%%%%%%%%%%%%%%%%%%%%%%%%%%%%%%%%

% Don't change these lines
\bsp	% typesetting comment
\label{lastpage}
\end{document}